\documentclass{jfm}
\usepackage{graphicx}
\usepackage{epstopdf, epsfig}
\usepackage{amsmath}
\usepackage{xcolor}
\usepackage{amssymb,upgreek}
\usepackage{lscape}
\usepackage{pdflscape}
\usepackage{graphicx}
\usepackage{subcaption}
\usepackage{float}
\usepackage{listings,amsmath,amssymb,xcolor,graphicx,url, upgreek}
\usepackage{esint}
\usepackage[mathscr]{eucal}
\usepackage{nomencl}
\usepackage[authormarkup=none]{changes}
\newcommand{\gobble}[1]{}
\setdeletedmarkup{\gobble{#1}}
\makeatother

\usepackage{xcolor}
\definecolor{dc}{rgb}{0.03, 0.27, 0.49}
\definecolor{pp}{rgb}{0.41, 0.16, 0.38}
\definecolor{dg}{rgb}{0.0, 0.44, 0.0}
\definecolor{jazzberryjam}{rgb}{0.65, 0.04, 0.37}
\definecolor{mygreen}{rgb}{0.0, 0.5, 0.0}
\definecolor{frenchrose}{rgb}{0.96, 0.29, 0.54}
\definecolor{brightpink}{rgb}{1.0, 0.0, 0.5}

\definechangesauthor[color=black]{R1}
\definechangesauthor[color=black]{R2}
\definechangesauthor[color=black]{R3}
\definechangesauthor[color=black]{ECJ}
\definechangesauthor[color=black]{RR2}
\definechangesauthor[color=black]{RR3}

\newcommand{\myparI}[2]{\frac{\partial #1}{\partial #2}}

\newcommand{\mydiff}[2]{\frac{\mathrm{d} #1}{\mathrm{d} #2}}

\newcommand{\uvec}[1]{\boldsymbol{#1}}
\newcommand{\erf}[0]{\mathrm{erf}}
\newcommand{\erfi}[0]{\mathrm{erfi}}

\shorttitle{The effect of ridges and grooves on static menisci in rectangular channels}

\title{The effect of isolated ridges and grooves on static menisci in rectangular channels}
\author{Eleanor C. Johnstone\aff{1}
	\corresp{\email{eleanor.johnstone@manchester.ac.uk}}, Andrew L. Hazel\aff{1} \and Oliver E. Jensen\aff{1}}

\affiliation{\aff{1}Department of Mathematics, University of Manchester, Manchester M13 9PL, UK}

\begin{document}

\maketitle

\begin{abstract}
We present theoretical and numerical results that demonstrate the sensitivity of the shape of a static meniscus in a rectangular channel to localised geometric perturbations in the form of narrow ridges and grooves imposed on the channel walls.  The Young--Laplace equation is solved for a gas/liquid interface with fixed contact angle using computations, analytical arguments and semi-analytical solutions of a linearised model for small-amplitude perturbations. 
We find that the local deformation of the meniscus's contact line near a ridge or groove is strongly dependent on the shape of the perturbation.  In particular, 
small-amplitude perturbations that change the channel volume \deleted[id=R3]{(evaluated where they intersect the contact line)}induce a change in the pressure difference across the meniscus, resulting in long-range curvature of its contact line.  
We derive an explicit expression for this induced pressure difference directly in terms of 
the boundary data. We 
show how contact lines can be engineered to assume prescribed patterns using suitable combinations of ridges and grooves. 
\end{abstract}


\section{Introduction}\label{sec:intro}
The behaviour of fluids when the dominant force is surface tension underpins many fundamental physical and industrially valuable processes: for example, microfluidics and inkjet printing \citep{calver2020thin, he2017roles, yang2005droplet}; directional transport of liquids in biological processes \citep{comanns2015directional, bhushan2019bioinspired, zheng2010directional, ju2012multi, prakash2008surface, xu2017}; engineering applications such \added[id=R3]{as} water harvesting \citep{li2017topological, xu2016high, brown2016bioinspired}; and the behaviour of fluids in low-gravity situations \citep{passerone2011twenty}. Moreover, from a purely theoretical point of view, such systems have been known to exhibit a plethora of complex behaviours associated with contact-line dynamics, such as contact-angle hysteresis \citep{gao2006contact, eral2013contact, dussan1979spreading} and the `stick-slip' phenomenon  \citep{shanahan1995simple, bourges1995influence, picknett1977evaporation, stauber2014lifetimes}. However, the behaviour we study here is that of a very simple static system, which forms the `basic state' for many of these dynamical problems. Specifically, we seek to quantify and describe the sensitivity of menisci in confined channels to imperfections in geometry.  Understanding such sensitivity is important in the industrial and biological processes described above, where natural or manufactured surfaces are in general not perfectly smooth.  Indeed, the sensitivity of microfluidic devices to small imperfections has hampered their usefulness in an industrial setting \citep{calver2020thin, zhou2012surface}. \added[id=R2]{On the other hand, the structuring of channels by parallel grooves has also been shown to improve the efficacy of microfluidic devices: for example, railed microfluidic channels have been used to create superhydrophobic surfaces \citep{emami2013predicting, yoshimitsu2002effects}, to guide and assemble microstructures inside fluidic channels \citep{chung2008guided}, and in primary cell culture technology to control the deposition and location of cells within microfluidic devices \citep{mobasseri2015polymer, khabiry2009cell, howell2005microfluidic, manbachi2008microcirculation, khademhosseini2004molded, khademhosseini2005cell, park2006microchannel, lee2007artificial}.}\added[id=RR2]{ \cite{stone2004engineering} also provide a general discussion of the role of channel geometry in controlling fluids in microchannels. \cite{anna2016droplets} and \cite{ajaev2006modeling} give a more general discussion of the modelling and applications of drops and bubbles in microchannels and confined channels.} 

It has been known since \cite{wenzel1936resistance} that menisci in channels are sensitive to imperfections in the channel geometry. Wenzel's work demonstrating the effect of wall roughness on contact-line wettability using simple energy-conservation arguments was further extended for porous surfaces by \cite{cassie1944wettability} and \cite{cassie1948contact}. Quasi-static effects such as contact-line hysteresis (at finite microscopic contact angle) and the stick-slip phenomenon were first observed by \cite{johnson1964contact} who studied the wettability of a drop on a rough surface where the roughness was in the form of concentric circular grooves. By moving the contact line very slowly over the obstacles, they observed multiple axisymmetric equilibria. \cite{huh1977roughness} studied surfaces with more complex roughness, including cross, hexagonal and radial grooves. They used the linearised Young--Laplace equation to find a relationship between the contact angle and hysteresis/stick-slip behaviour. Surfaces with periodic roughness \citep{cox1983spreading} and random roughness \citep{jansons1985moving} were also found to induce contact-angle hysteresis and stick-slip behaviour in the limit of zero capillary number. \added[id=R1]{More recently, rough surfaces with Gaussian-type defects have been shown to significantly influence contact line dynamics in the contexts of droplet spreading \citep{espin2015droplet}, droplet sliding \citep{park2017droplet}, and droplet evaporation and imbibition \citep{pham2017drying, pham2019imbibition}}. \cite{jansons1985moving} made the further observation that the location of the contact line influenced its future position, leading to irreversibility of the wetting process. 
Stick-slip behaviour can also be induced by defects caused by changes in wettability or temperature \citep{ajaev2020interaction}.

\cite{concus1969behavior}, \cite{fowkes1998surface} and \cite{reyssat2014drops} showed that for static liquid-vapour interfaces in a wedge, the existence of solutions depends on the angle of the wedge and the contact angle between the liquid-vapour and solid-liquid interface. Instead of imposing perturbations on the channel walls geometrically, it is also possible to perturb the meniscus by changing the contact angle locally on the upper and lower channel walls. \citet{boruvka1978analytical} considered a meniscus in contact with a strip-wise heterogeneous wall in which each strip has a different equilibrium contact angle. They showed that locally near the wall, the contact-line \replaced[id=R3]{curvature}{displacement} becomes unbounded at the point where the contact angle changes. The jump in contact angle is analogous to a ridge or groove perturbation with corners; here the \replaced[id=R3]{displacement of the contact line can be unbounded in some circumstances}{solution on the wall} \citep{concus1969behavior, weislogel1996capillary, king1999laplace}. In what follows we consider smooth (differentiable everywhere) wall perturbations with no sharp corners.

At low flow speeds, the motion of drops and bubbles in confined devices follows a series of near-equilibrium configurations and thus the static problem discussed here can also be used to provide insight into the effect of wall roughness on these problems. Channel imperfections are known to have a significant effect on the set of observable stable solutions for air fingers and bubbles propagating in a Hele-Shaw channel. This effect has been seen with a cusp at the tip of a bubble/finger created by a needle \citep{hong1988bubbles}; a tiny bubble at the tip of a bubble/finger \citep{maxworthy1986bubble}; anisotropy by etching of the plates \citep{ben1985experimental, dorsey_martin_1987}; and channel occlusions \citep{hazel2013multiple,thompson2014multiple}. Wall roughness has also been seen to affect the `tip-splitting' effect seen by propagating interfaces \citep{tabeling1987experimental, franco2016sensitivity, franco2018bubble}. It is so far unclear how the wall roughness contributes to the selection of a particular set of solutions. 

In this study, we consider a static liquid-vapour interface in a rectangular channel and introduce imperfections to the upper and lower walls in the form of narrow ridges and grooves running the length of the channel. We are interested in how the meniscus shape changes due to the perturbations and, in particular, how the perturbations displace the contact line (defined as the intersection of the meniscus with the channel walls). We consider two classes of perturbations: those that change the volume of the channel, and those that preserve it. A key result is that small-amplitude perturbations that change the volume of the channel induce a change in the pressure difference across the meniscus, and thus change the mean curvature of the meniscus. This change is a long-range effect that is felt along the whole contact line. 

In \S\ref{sec:model} we present the governing nonlinear Young--Laplace equation and boundary conditions.  In \S\ref{sec:modellin} we derive and solve a linear model to find the shape of the meniscus for perturbations of small amplitude. The linearised problem shows that the change in mean curvature of the meniscus (\hbox{i.e.} the change in pressure difference over the meniscus) due to small perturbations is given by the Helmholtz equation. We derive an explicit expression for the change in pressure difference which is directly proportional to the integral of the perturbations around the contact line, which corresponds to the change in channel volume.  We also solve for the fully nonlinear mean curvature of the meniscus numerically using Surface Evolver \citep{evolver}.

We present results for both the linear and nonlinear models in \S\ref{sec:results}. We show that for channel-volume-changing perturbations, the meniscus away from the local perturbation matches onto a catenoid which must have the same constant mean curvature as the meniscus, which can be worked out \textit{a priori} from the boundary data.  \replaced[id=R3]{We also show that perturbations that are offset from each other on the two walls, for example forming weakly-corrguated channels, can be used to engineer patterns in the contact line because each perturbation causes a deflection of both the upper and lower contact lines. The deflection mechanism acts differently depending on whether the total perturbation is channel-volume changing or channel-volume preserving.}{We also show that perturbations that are offset from each other on the two walls can be used to engineer patterns in the contact line, either through a \replaced[id=R3]{deflection}{scattering} mechanism (for channel-volume-preserving perturbations) or by exploiting the long-range curvature induced by a pressure change.}  We conclude with a short discussion in \S\ref{sec:conclusion}.
\section{Model}\label{sec:model}

	\begin{figure}
	\centering
	\includegraphics[width=1\textwidth]{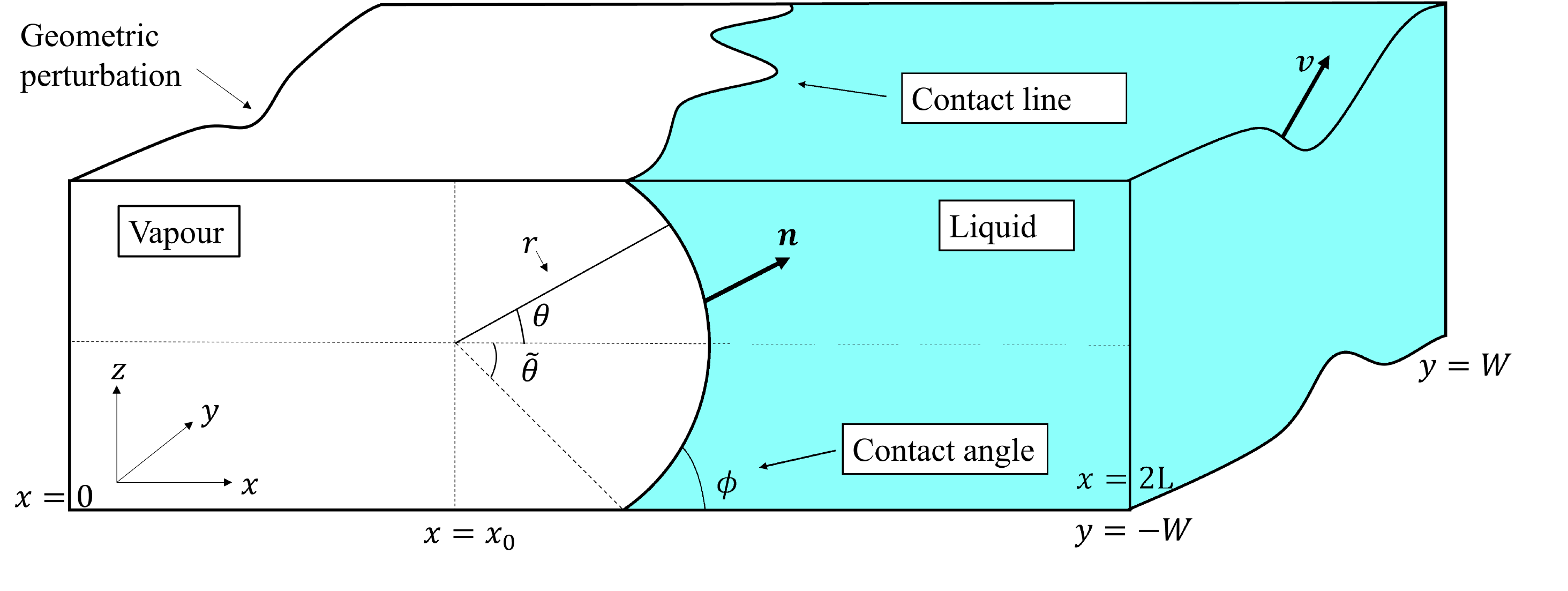}
	\caption{A static liquid-vapour meniscus in a rectangular channel $0 \leq x \leq 2L$, $-W \leq y \leq W$ and $-\tfrac 1 2 + B_-(y) \leq z \leq \tfrac 1 2 + B_+(y)$. The shape of the meniscus is described using a cylindrical polar coordinate system $(r, \theta, y)$ centred at $(x, y, z) = (x_0, 0, 0)$ where $x_0$ is fixed by the volume of liquid in the channel. The meniscus is located in the channel such that the unperturbed state has $r=R=1/(2\cos \phi)$, with $\phi$ the solid-liquid contact angle on the upper and lower walls. \added[id=R3]{The solid-liquid contact angle on the side walls is $\upi/2$.}}
	\label{fig:1}
\end{figure}

Consider a static liquid-vapour interface having uniform mean curvature in a rectangular channel with non-dimensional edge lengths $2L$, $2W$ and $1$ in the $x$, $y$ and $z$ directions respectively; see figure~\ref{fig:1}. \added[id=R2]{We assume that the channel is sufficiently small that gravitational effects can be ignored, \hbox{i.e.} that the height of the horizontal channel is much smaller than the capillary lengthscale $\sqrt{\gamma_{LV}/\rho g}$, where $\gamma_{LV}$ is the liquid-vapour surface tension, $\rho$ the liquid density and $g$ the acceleration due to gravity.}
The contact angle between the liquid-vapour and solid-liquid interface on the upper and lower walls of the channel is fixed to be $\phi$, where $0 \leq \phi < \pi/2$. We impose that the meniscus meets the side walls at $y = \pm W$ normally, 
with contact angle $\pi/2$,
so that in the absence of wall perturbations, the interface takes the shape of the arc of a cylinder.  To describe this, we adopt cylindrical polar coordinates $(r, \theta, y)$ with $r=0$ fixed at the centre of curvature of the unperturbed meniscus. We define the maximum value of $\theta$, which specifies the contact-line location in the unperturbed state, by $\added[id=R2]{\added[id=R2]{\tilde \theta}} = \pi/2 - \phi$ (see figure~\ref{fig:1}). Then the base state is given by $r= R \equiv 1/(2\sin\added[id=R2]{\added[id=R2]{\tilde \theta}})$, for $\theta \in [-\added[id=R2]{\tilde \theta},\  \added[id=R2]{\tilde \theta}\ ]$ and $y \in [-W,\ W]$. Cartesian and polar coordinates are related by $(x, y, z) = (x_0 + r \cos\theta, \ y, \ r\sin\theta)$, where $x_0$ is defined by the volume of liquid $V_L$ in the channel, as
\begin{equation}
x_0 = 2L - \frac{R}{2}\cos\added[id=R2]{\tilde \theta} -  \frac{V_L}{2W}  -R^2  \added[id=R2]{\tilde \theta}.
\end{equation} 
We denote the dimensionless pressure of the liquid phase (scaled on surface tension over channel depth) to be $p_L$, relative to zero pressure in the gas phase, 
so that in the unperturbed configuration $p_L=-1/R$. 

The upper ($+$) and lower ($-$) walls of the channel are then perturbed so that they are described by $z = \pm \tfrac 1 2 + B_{\pm}(y)$. The perturbations take the form of ridges and grooves which could be created, for example, \replaced[id=R2]{using a pulsed laser \citep{xing2014multiple}, or using moulded fabrication with standard soft lithography techniques \citep{chung2008guided}.}{by adding a strip of paint or etching the channel walls with a needle.} \added[id=R2]{We do not assume contact angle hysteresis, but we assume that surfaces are homogeneous and smooth, allowing us to address interactions between the wall perturbations and the meniscus at the microscopic level.}
We non-dimensionalise all surface \replaced[id=ECJ]{tensions}{energies} on the liquid-vapour surface tension $\gamma_{LV}$ so that the meniscus has unit surface tension.

We specify the interface location relative to the base state so that the surface of the meniscus is described in terms of the unknown radial perturbation $F(y, \theta)$ and the angular change in location of the contact lines on the upper and lower walls $\Phi_\pm(y)$:
\begin{equation}\label{eq:nlintshape}
r = R + F(y, \theta), \quad \theta \in [ -\added[id=R2]{\tilde \theta} + \Phi_-(y), \ \added[id=R2]{\tilde \theta} + \Phi_+(y)], \quad y \in [-W, W].
\end{equation}
 \added[id=ECJ]{We measure the pressure difference across the meniscus as the liquid pressure minus the gas pressure, where without loss of generality we assume the gas pressure is zero.} Thus we write $\Delta p = -R^{-1} + (p_L)_p$, where $(p_L)_p$ is the change in pressure of the liquid phase due to the channel perturbations. This is constrained by the requirement that the volume of liquid $V_L$ does not change under any perturbation to the channel geometry. Then, defining the unit normal $\uvec n$ of the meniscus to point into the liquid phase, as shown in figure~\ref{fig:1}, the equilibrium state is specified by the
Young--Laplace equation, relating the uniform mean curvature of the interface to the pressure difference across the meniscus $\Delta p$ as
\begin{equation}\label{eq:YoungLap3d}
\Delta p = -\nabla \cdot \uvec n\vert_{r=R+F}=
\added[id=R3]{
-\frac{1}{\Lambda} + \left(\frac{(R+F)F_y}{\Lambda}\right)_y+\frac{1}{R+F}\left(\frac{F_\theta}{\Lambda}\right)_\theta,}
\end{equation} 
where
\added[id=R3]{
$\Lambda \equiv\sqrt{(R+F)^2(1+F_y^2)+F_\theta^2}$.
}

We solve the Young--Laplace equation \eqref{eq:YoungLap3d} subject to the volume constraint and the following boundary conditions. First, we constrain the contact line to lie on the perturbed channel walls, so that
\begin{equation}\label{eq:nlbc1}
z = \Big(R + F(y, \theta) \Big) \sin\theta = \pm \tfrac 1 2  + B_{\pm}(y) \quad \mbox{at} \enspace \theta = \pm \added[id=R2]{\tilde \theta} + \Phi_{\pm}(y).
\end{equation}
Second, we impose a fixed contact angle $\phi$ through the geometrical argument that if $\uvec{v}_{\pm}$ are unit normals to the upper and lower channel walls pointing out of the channel (as shown in figure~\ref{fig:1}), then
\begin{equation}\label{eq:nlbc2}
\uvec n \cdot \uvec{v}_{\pm} = \cos\phi \ \mbox{at} \ \theta = \pm \added[id=R2]{\tilde \theta} + \Phi_{\pm}(y), \quad \mbox{where} \quad \uvec{v}_{\pm} = \pm \frac{- B_{\pm}'(y)\uvec y+ \uvec z }{\sqrt{1 + B_{\pm}'(y)^2}}.
\end{equation}
Finally, the meniscus meets the side walls normally, so that
\begin{equation}\label{eq:nlbc3}
F_y(\pm W, \theta) = 0.
\end{equation}

\subsection{Linear model}\label{sec:modellin}
We linearise the problem when wall perturbations are small relative to the channel depth by writing $B_{\pm}(y) = \epsilon b_{\pm}(y)$, where $\epsilon$ is defined as the maximum amplitude of the perturbation and $b_{\pm}(y) = O(1)$ as $\epsilon \to 0$. We use the parametrisation of the interface given in the nonlinear model \eqref{eq:nlintshape}, with the assumption that the perturbation to the radius and the change in location of the contact lines are also $O(\epsilon)$, so that $F(y, \theta) = \epsilon f(y, \theta)$ and $\Phi_\pm(y) = \epsilon \Theta_\pm(y)$. Then the interface is parametrised as 
\begin{align}
r = R + \epsilon f(y, \theta), \quad \theta \in \left[- \added[id=R2]{\tilde \theta}+ \epsilon \, \Theta_-(y), \ \added[id=R2]{\tilde \theta} +\epsilon \, \Theta_+(y) \right], \quad y \in [-W,\,  W].
\end{align} 
We assume that the change in the pressure difference due to the perturbation is also $O(\epsilon)$, so that $(p_L)_p = \epsilon p$ and 
$\Delta p = -R^{-1} + \epsilon p$.\\

Assuming $f(y, \theta)$ and all its derivatives are $O(1)$ as $\epsilon \to 0$, then after linearisation, the leading-order approximation to the Young--Laplace equation \eqref{eq:YoungLap3d} is the Helmholtz equation
\begin{equation}\label{eq:helmholtz}
\frac{1}{R^2}f +\frac{1}{R^2}f_{\theta\theta} + f_{yy} =p, \quad y \in [-W, W], \ \theta \in [-\added[id=R2]{\tilde \theta}, \added[id=R2]{\tilde \theta}].
\end{equation}
The boundary condition \eqref{eq:nlbc1} constraining the contact line to lie on the upper and lower channel walls becomes
\begin{equation}\label{eq:lbc1}
\replaced[id=R3]
{f(y, \pm \added[id=R2]{\tilde  \theta})\sin\added[id=R2]{\tilde \theta} \pm f_\theta(y,\pm \added[id=R2]{\tilde \theta}) \cos\added[id=R2]{\tilde \theta} = \pm b_\pm(y)}
{f(y, \pm \tilde  \phi)\sin\added[id=R2]{\tilde \theta} =\pm \Big( b_\pm(y) -  f_\theta(y,\pm \added[id=R2]{\tilde \theta}) \cos\added[id=R2]{\tilde \theta} \Big)},
\end{equation}
whilst the leading-order approximation to boundary condition \eqref{eq:nlbc2} is
\begin{equation}\label{eq:lbc2}
\replaced[id=R3]
{f_\theta(y, \pm \added[id=R2]{\tilde \theta}) - R \, \Theta_\pm (y) = 0.}
{\Big(f_\theta(y, \pm \added[id=R2]{\tilde \theta}) - R \, \Theta_\pm (y)\Big)\cos\added[id=R2]{\tilde \theta} = 0.}
\end{equation}
In deriving \eqref{eq:lbc2}, the neglected $O(\epsilon^2)$ terms include those involving the derivative of the boundary data, $\epsilon^2 b_{\pm}'(y)$. These terms will formally become $O(\epsilon)$ if $b_{\pm}'(y) = O(\epsilon^{-1})$; this puts an additional constraint on the boundary data for the linearised model to be valid. In particular, this constraint suggests that we cannot use a linearised model for very sharp perturbations with large gradients regardless of their amplitude; this will be discussed in \S\ref{sec:methgauss}. 
The final boundary condition \eqref{eq:nlbc3} 
becomes
\begin{equation}\label{eq:lbc3}
f_y(\pm W, \theta) = 0.
\end{equation}
The problem is closed with the condition that the volume of liquid $V_L$ is invariant \added[id=R3]{with respect to changes in channel volume}. \replaced[id=R3]{This}{ 
After integrating to find the volume of the perturbed channel $V_C$ and the volume of vapour in the perturbed channel $V_V$, the condition $V_C = V_V + V_L$ for any geometry} leads to \added[id=R3]{the condition}  
\begin{align}\label{eq:volume}
\int_{-W}^{W} \int_{-\added[id=R2]{\tilde \theta}}^{\added[id=R2]{\tilde \theta}} f(y, \theta) \ \mathrm d \theta \ \mathrm d y = \left(\frac{V_L}{2RW} + \frac{\added[id=R2]{\tilde \theta}}{2\sin\added[id=R2]{\tilde \theta}} -\frac{\cos\added[id=R2]{\tilde \theta}}{2}\right) \int_{-W}^{W} \Big( b_+(y) - b_-(y)\Big) \ \mathrm d y. 
\end{align}
Finally, the linearised contact-line displacement on the upper and lower walls $x_\pm(y)$ is found by Taylor expanding the solution for $x = r \cos\theta$ at $\theta = \pm \added[id=R2]{\tilde \theta} + \epsilon \Theta_\pm(y)$:
\begin{align}\label{eq:cleqn}
x_\pm(y)  =  x_0 + R \cos\added[id=R2]{\tilde \theta}  + \epsilon x_p^\pm(y)  + O(\epsilon^2), \quad
x_p^\pm(y) =  f(y, \pm \added[id=R2]{\tilde \theta} )\cos\added[id=R2]{\tilde \theta} \mp f_\theta(y, \pm \added[id=R2]{\tilde \theta}) \sin\added[id=R2]{\tilde \theta}.
\end{align}

For the specific case of zero contact angle ($\phi=0$), when the meniscus meets the walls tangentially, \replaced[id=R3]{there is no contribution to boundary condition \eqref{eq:lbc2} at $O(\epsilon)$. An expansion to powers of $O(\epsilon^2)$ is needed to obtain \eqref{eq:lbc2}, }{\eqref{eq:lbc2} disappears because the normal to the meniscus $\uvec n$ is zero at $O(\epsilon)$. An expansion to powers of $O(\epsilon^2)$ is needed to obtain the boundary condition; however, the problem remains the same}as explained in Appendix~\ref{app:zerocont}.

\section{Methods}\label{sec:methods}
\subsection{Pressure-volume relationship}
We show that small-amplitude perturbations that change the volume of the channel \deleted[id=R3]{(as evaluated at the contact lines)}must cause the mean curvature of the meniscus to change. Noting the self-adjointness of the Helmholtz operator and boundary conditions (\ref{eq:helmholtz})--\eqref{eq:lbc3}, we derive in Appendix~\ref{app:indpressure} an explicit expression for this $O(\epsilon)$ pressure difference at the meniscus and find that it has a similar dependence on the boundary data as the induced volume change, 
\begin{equation}\label{eq:indpressure}
p =  \frac{1}{4 W R^2 \sin (\added[id=R2]{\tilde \theta})} \int_{-W}^{W} \Big( b_+(y) - b_-(y)\Big) \ \mathrm d y.
\end{equation}
Thus in the linear problem with channel-volume-changing perturbations (for which $\int_{-W}^{W} ( b_+(y) - b_-(y)) \ \mathrm d y \neq 0$), the forcing for the Helmholtz equation (\ref{eq:helmholtz}) can be deduced \textit{a priori} from the volume change encoded in the boundary data.

\subsection{Solutions for Gaussian perturbations}\label{sec:methgauss}
We now restrict our attention to Gaussian perturbations of the form
\begin{equation}
\label{eq:gausspert}
\added[id=R3]{B_\pm(y) =  a^\pm }\epsilon \ \mathrm e^{-(y-y_c^\pm)^2/s^2}, \quad \mbox{so}\quad \ b_\pm(y) = a^\pm \mathrm e^{-(y-y_c^\pm)^2/s^2},
\end{equation}
where $y_c^\pm$ and $s$ control the location and width of the perturbation respectively and \added[id=R3]{the prefactor $a^\pm$, which can take the value $+1$ or $-1$ on either wall,} determines the orientation of the perturbation, i.e. whether it is a ridge or a groove. For the purposes of illustration, we assume that the perturbation on the lower wall is a ridge so that \replaced[id=R3]{$a^-=1$}{$B_-$ has positive amplitude}. Then after fixing $s$, we consider two specific types of geometry: channel-volume-preserving configurations with $\int_{-W}^{W} ( b_+(y) - b_-(y)) \ \mathrm d y = 0$ (corresponding to a groove on the upper wall); and channel-volume-changing configurations with $\int_{-W}^{W} ( b_+(y) - b_-(y)) \ \mathrm d y  < 0$ (corresponding to a ridge on the upper wall). The former 
preserve the pressure of the liquid phase, whereas the latter, 
which decrease the volume of the channel, cause an increase to the magnitude of the liquid pressure. If $y_c^\pm=0$ then the channel-volume-preserving and channel-volume-changing configurations are mirror-anti-symmetric and mirror-symmetric respectively about $z=0$. 

We solve the full nonlinear problem to find \added[id=R3]{the} shape of the interface using the open-source software Surface Evolver \citep{evolver}, which uses a gradient-descent method to converge to a surface with minimum energy from a given initial guess and subject to constraints \added[id=ECJ]{to enforce the boundary conditions and the volume condition}; for details see Appendix~\ref{app:evolverexpl}. Meanwhile, we solve the linear problem (\ref{eq:helmholtz})-(\ref{eq:volume}) using second-order-accurate finite differences; see Appendix~\ref{app:linmethods} for more details. As seen in \S\ref{sec:modellin}, the linear model can be expected to break down when $b_\pm'(y) = O(\epsilon^{-1})$. For the Gaussian boundary data \eqref{eq:gausspert}, this occurs if $s^2 =O(\epsilon)$, which limits how narrow the perturbation can be.

\subsection{Analytic solution for symmetric perturbations }\label{sec:analyticsol}
For the special case of aligned perturbations ($y_c^\pm=0$), we can obtain an analytic solution of the linear problem (\ref{eq:helmholtz})-(\ref{eq:volume}) via separation of variables, with a Fourier discretisation across the width of the channel in the $y$ direction because of the $y$-dependence of the boundary conditions on the upper and lower walls. Denoting mirror anti-symmetric (channel-volume-preserving) and mirror-symmetric (channel-volume-changing) solutions by `MAS' and `MS' subscripts respectively, the series solution is given by
\begin{equation}\label{eq:seriesol}
 f_{\text{MAS}\atop\text{MS}}(y, \theta)   = R^2 p+ \sum_{n=0}^\infty A_n^{\text{MAS}\atop\text{MS}} \left(   \mathrm e^{\lambda_n\theta  } \mp \mathrm e^{-\lambda_n\theta } \right) \cos\left(\frac {n\pi y }{W}\right),
\end{equation}
where the exponential coefficient $\lambda_n$ is given by
\begin{equation}\label{eq:lamn}
\lambda_n = \sqrt{\left(\frac{n\pi R}{W}\right)^2 -1}.
\end{equation}
Thus there is a critical value of $n$, specific to the contact angle (through $2R \cos\phi=1$) and the width of the channel, at which the sum switches from having oscillatory dependence in $\theta$ to exponential dependence in $\theta$. The coefficients $A_n^{\text{MAS}\atop\text{MS}} $ are given by
\begin{align}\label{eq:Anpm}
&A_n^{\text{MAS}\atop\text{MS}} =  \frac {\pm a_n}{ \left(\sin \added[id=R2]{\tilde \theta}  +  \lambda_n\cos  \added[id=R2]{\tilde \theta} \right)  {{\rm e}^{\lambda_n\added[id=R2]{\tilde \theta}}} \mp   \left( \sin \added[id=R2]{\tilde \theta}   -\lambda_n\cos  \added[id=R2]{\tilde \theta}  \right) {{\rm e}^{-\lambda_n\added[id=R2]{\tilde \theta}}}} \quad \begin{cases} n \geq 0 \ (\text{MAS}), \\ n \geq 1 \ (\text{MS}), \end{cases} \\
&a_0^{\text{MAS}} = \frac {1}{2W} \int_0^{W} b_-(y) \ \mathrm d y, \quad a_n = \frac {1}{W} \int_0^{W} b_-(y)  \cos\left(\frac {n\pi y }{W}\right) \ \mathrm d y, \quad n \geq 1.
\end{align}
The coefficient $A_0^{\text{MS}}$ is  found using the volume condition \eqref{eq:volume} to be
\begin{equation}
A_0^{\text{MS}} =  \frac{\left(\dfrac{V_L}{2RW} + \dfrac{\added[id=R2]{\tilde \theta}}{2\sin\added[id=R2]{\tilde \theta}} -\dfrac{\cos\added[id=R2]{\tilde \theta}}{2}\right) \displaystyle \int_{-W}^{W} \Big( b_+(y) - b_-(y)\Big) \ \mathrm d y - 4 W R^2  \added[id=R2]{\tilde \theta} \ p}{4 W \sin \added[id=R2]{\tilde \theta}}.
\end{equation}
For the Gaussian boundary data \eqref{eq:gausspert}, the coefficients of the convergent series are defined by
\begin{equation}\label{eq:coeffBN}
a_n  = \frac {s\sqrt{ \pi}} {2W} \exp\left(-\frac {s^2} 4 \left(\frac {n\pi }{W} \right)^2\right)\mbox{Re} \left\{ \mathrm i  \mathrm{erfi} \left( -\frac{\mathrm i W}{s} - \frac {n \pi s}{2W} \right)\right\}, \quad n \geq 1.
\end{equation}
The function $\erfi(z) = -\mathrm i \erf(\mathrm i z)$ is the imaginary error function so that for real $u$ and $v$,  $\mathrm i\erfi(-\mathrm i u+v) = \erf(u+\mathrm i v
)$.

Numerical solutions below are obtained by truncating \eqref{eq:seriesol} at $n=n_c$ such that \replaced[id=ECJ]{terms with coefficients smaller than $10^{-16}$ were discarded.}{the coefficients $A_{n_c}^{\text{MAS}/\text{MS}} \sim 10^{-16}$.}


\section{Results}\label{sec:results}
We present results for the displacement of the static meniscus and the contact line induced by Gaussian perturbations \eqref{eq:gausspert}. We first assume that the perturbations are aligned so that $y_c^\pm = 0$.

\subsection{Aligned perturbations}\label{sec:symresults}

\begin{figure}
	\centering
		(a)\includegraphics[width=1.03\textwidth]{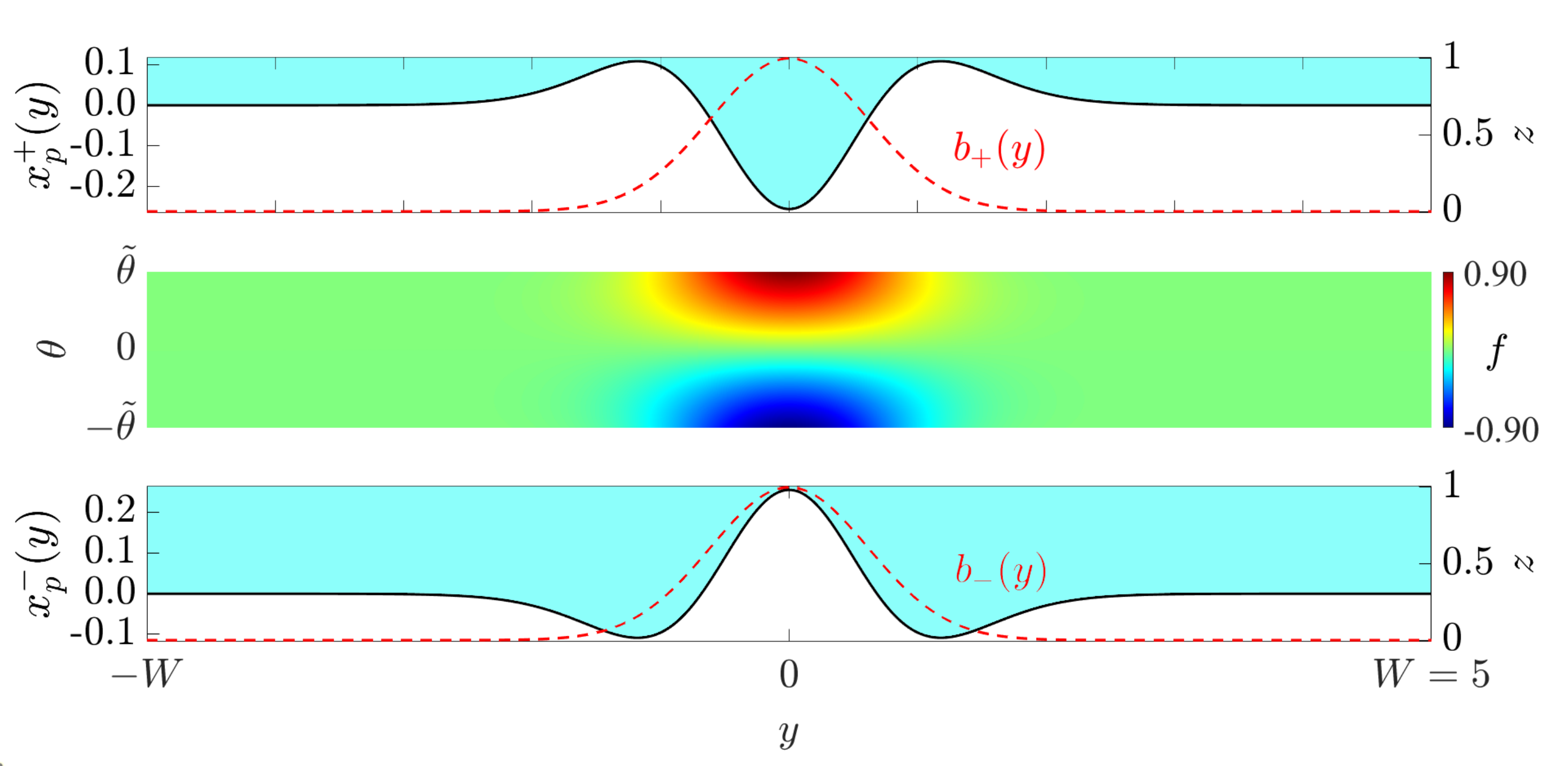}\label{fig:1a}
		(b)\includegraphics[width=1.03\textwidth]{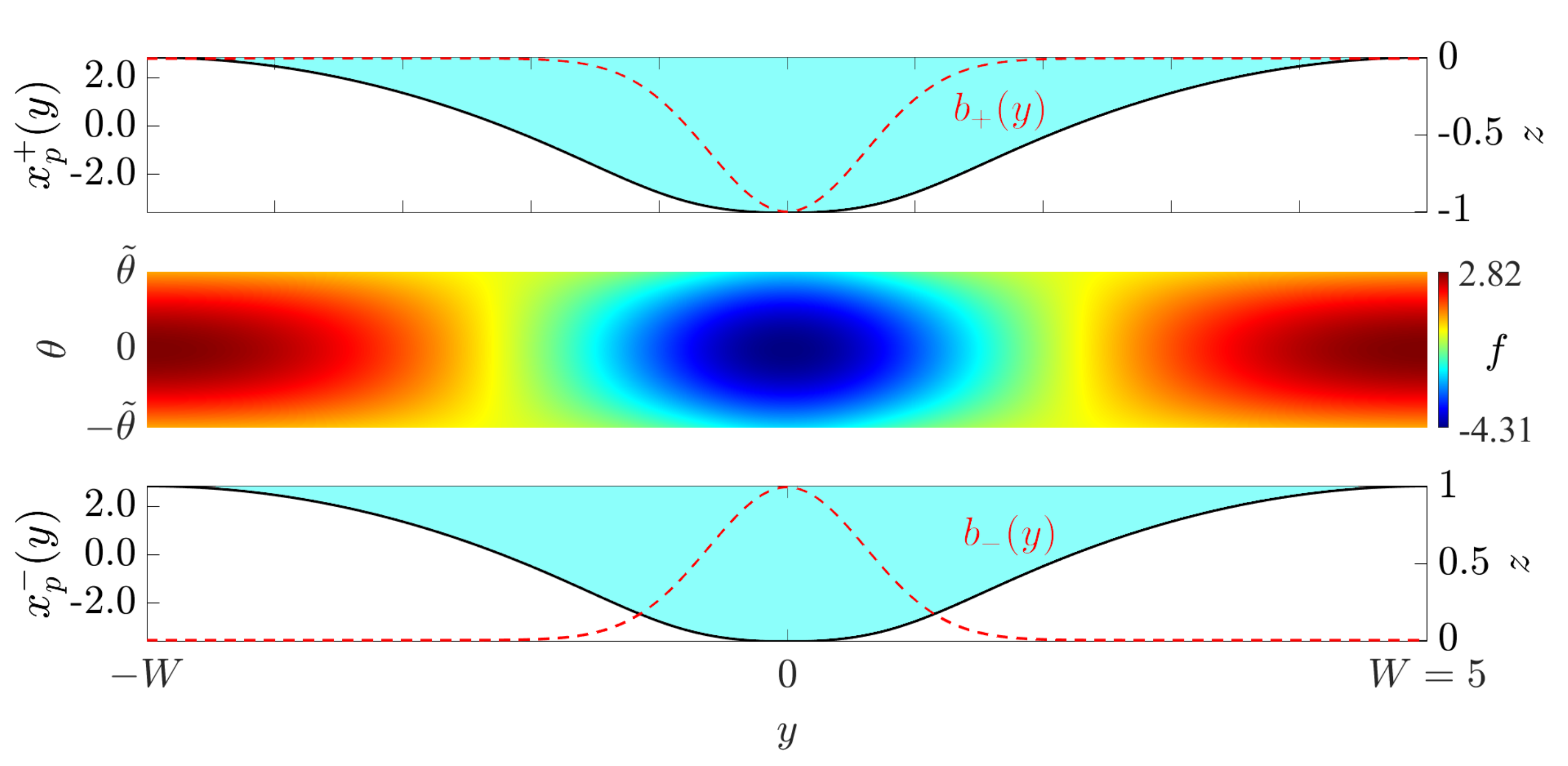}\label{fig:1b}
	\caption{The displacement of the upper and lower contact lines due to the perturbation $x_p^\pm(y)$ (black solid lines, left axis) and change in meniscus shape $f(y, \theta)$ (heat map) due to Gaussian perturbations $b_\pm(y)$ proportional to $ b(y) = \exp(-y^2/0.75)$ (red dashes, right axis), for a channel with half-width $W = 5$ and contact angle $\phi = 15^{\circ}$. As in figure~\ref{fig:1}, the liquid and vapour sides of the contact-line displacement are shown with blue and white shading respectively for: (a) channel-volume-preserving perturbation ($b_- = b_+ = b$); (b) channel-volume-changing perturbation ($b_- = -b_+ = b$). The heat maps denote the change in radius $r$ due to the perturbations, with green denoting no change in the meniscus location. Positive values of $f$ show the meniscus extending into the liquid phase.}
	\label{fig:fig6}
\end{figure}

Figure~\ref{fig:fig6} shows `baseline' linear solutions $f(y, \theta)$ for the two prototype channel-volume-preserving and changing configurations, together with displacement of the upper and lower contact lines due to the perturbation, $x_p^\pm(y)$ (as given in \eqref{eq:cleqn}), computed using the series solution \eqref{eq:seriesol}. 
In the channel-volume-preserving case (figure~\ref{fig:fig6}a), the response of the meniscus and contact line is localised around the perturbations \added[id=R3]{in the $y$ direction}, whereas in the channel-volume-changing case (figure~\ref{fig:fig6}b) the perturbations induce a larger-amplitude response that is felt across the entire depth and width of the channel. In the former case, the contact-line shape appears to mirror the curvature of the wall perturbation, but is smoother in the latter case.  
Thus a small ridge or groove placed in the centre of the channel can cause non-local bending of the contact line through its impact on the pressure field (\ref{eq:indpressure}). \added[id=R3]{Note that we can build new small-amplitude solutions as linear combinations of the two solutions shown, and can therefore describe the behaviour of the contact line as the perturbations are varied between the two configurations.
}

\begin{figure}
	\centering
	(a)\includegraphics[width=1.03\textwidth]{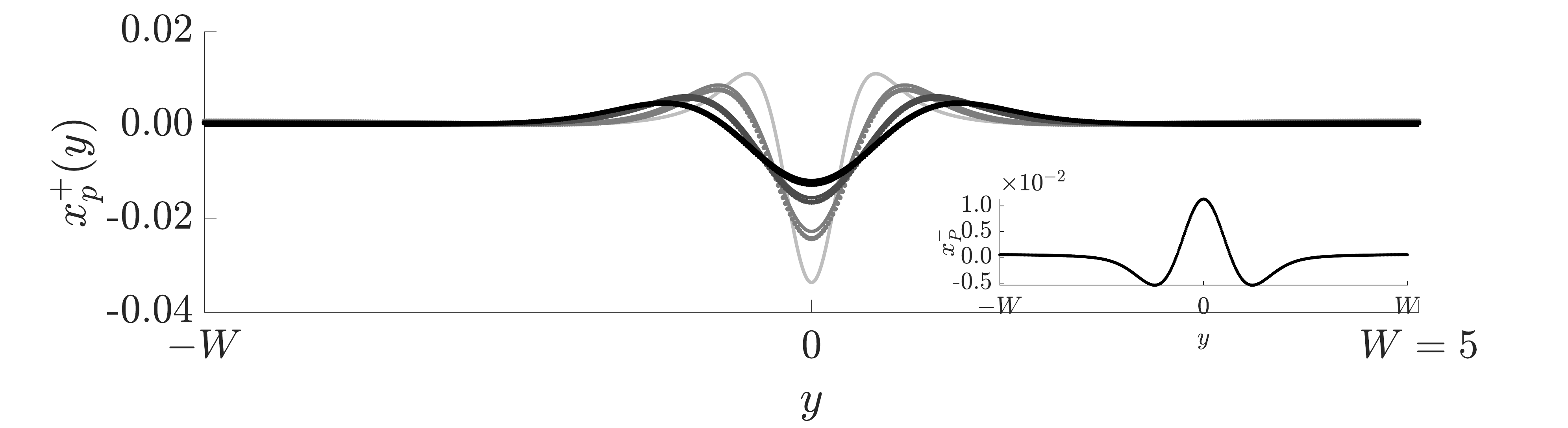}	\label{fig:2a}
	(b)\includegraphics[width=1.03\textwidth]{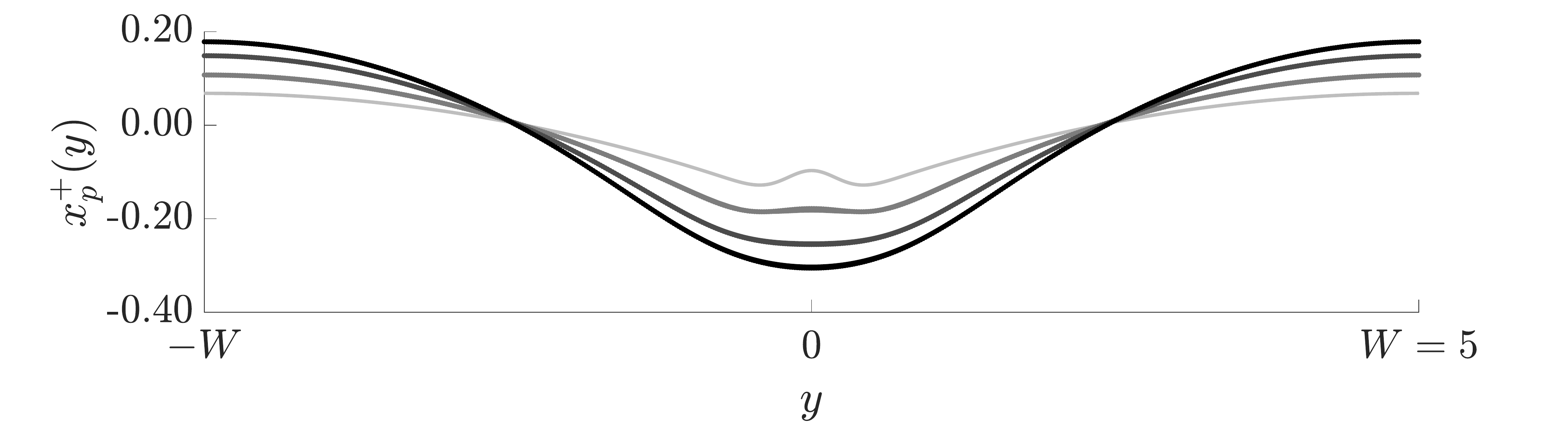}
	\label{fig:2b}
	(c)\includegraphics[width=1.03\textwidth]{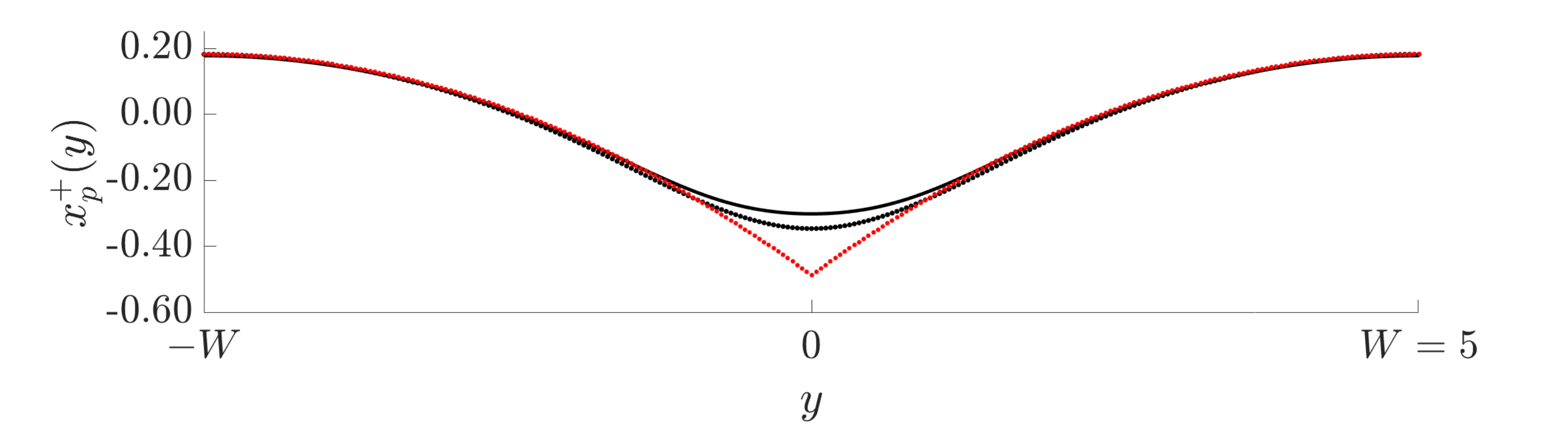}
	\label{fig:2d}
	\caption{The displacement of the upper contact line in channels of half-width $W=5$, for perturbations with amplitude $\epsilon = 0.01$ (a, b) and $\epsilon =0.1$ (c), for channel-volume-preserving perturbations (a) and channel-volume-changing perturbations (b, c). The contact angle is $\phi = 85^{\circ}$. (a, b) show wall perturbations of varying width, with darker colours indicating wider perturbations, with $s^2$ varying from $0.75$ (black), $0.5$, $0.25$ to $0.1$ (light gray). (c) shows $s^2 = 0.75$ only.  Solid lines denote the displacement calculated via the linear solution, \hbox{i.e.} $x_p^+(y)$ from \eqref{eq:cleqn}, for all values of $s$; circles \added[id=R2]{(thicker lines)} denote the nonlinear displacement $(x_{\text{cl}}-x_0-R\cos \added[id=R2]{\tilde \theta})/\epsilon$ where $x_{\text{cl}}$ is the contact-line data computed in Surface Evolver for $s^2 \geq 0.25$. Inset: the nonlinear lower contact-line displacement for volume-preserving perturbations, for $s^2=0.75$. The red dots in (c) denote the quadratic linear far-field solution \eqref{eq:farfieldsol}, with $C \approx 0.18$. }
	\label{fig:f2}
\end{figure}

The effect of varying perturbation amplitude and width on the contact-line displacement is presented in figure~\ref{fig:f2}.
Both volume-preserving 
(figure~\ref{fig:f2}a) and 
volume-changing 
(figure~\ref{fig:f2}b,c) perturbations result in a local displacement of the contact line in the centre of the channel which depends on the shape of the perturbations, with narrower or larger-amplitude perturbations eliciting a greater displacement.  Linear and nonlinear predictions of the upper contact-line displacement $x_p^+(y)$ show good agreement for perturbations which are approximately $1\%$ of the channel depth (figure~\ref{fig:f2}a, b). \added[id=R2]{(Numerical evidence in figure~\ref{fig:appf1} below suggests that the contact-line displacement increases like $\log(1/s)$ for finite contact angle as the perturbation width $s$ becomes very small, when the linear model breaks down.)} Larger perturbations, of approximately $10\%$ of channel depth, are shown in figure~\ref{fig:f2}(c) for the volume-changing case. 
Here a greater discrepancy between linear and nonlinear predictions is evident, 
although 
the shape of the contact-line 
perturbations at the edges of the channel 
is accurately described by the linear model. \replaced[id=RR3]{A long-wave analysis for $s \ll W$ shows that the solution in the far-field has}{An analysis of the linear solution \eqref{eq:seriesol} near the side-walls of the channel, where $y/W \sim 1$ \added[id=R3]{(for fixed $W=O(1)$ and $s\ll W$)}, suggests} a quadratic dependence on $y$ \added[id=R3]{in the form }\replaced[id=RR3]{$f_A(y, \theta) = C_1(\theta)(W-|y|)^2+C_2(\theta) $}{$(W-|y|)^2$}. \replaced[id=R3]{We substitute}{Substituting} this ansatz into the Helmholtz equation \eqref{eq:helmholtz} \added[id=R3]{and solve the resulting coupled ODEs for $C_1(\theta)$ and $C_2(\theta)$} with \replaced[id=R3]{modified `far-field'}{far-field} boundary conditions\replaced[id=R3]{. These are obtained by setting $b_\pm(y) = 0$ in \eqref{eq:lbc1} since the amplitude of Gaussian perturbations $b_\pm(y)$ is negligibly small for $y \sim W$ and $\vert y-y_c\vert \gg s$.}{(setting $b_\pm(y)= 0$)} \added[id=R3]{Then, substituting the solution for $f_A(y, \theta)$ in \eqref{eq:cleqn}} shows that the \added[id=R3]{approximation to the} displacement of the contact line \added[id=R3]{far from} the perturbations is given by
\begin{equation}\label{eq:farfieldsol}
\hat x_p^+ (y; \added[id=R2]{\tilde \theta}) \approx  \left({\frac {p\,\sin \left( \added[id=R2]{\tilde \theta} \right) }{\cos \left( \added[id=R2]{\tilde \theta}
		\right) \sin \left( \added[id=R2]{\tilde \theta} \right) +\added[id=R2]{\tilde \theta}}}\right)(W-|y|)^2 + C,
\end{equation}
where the translational constant $C$ cannot be found using the volume condition and instead is found empirically by \replaced[id=R3]{comparison with the value of the}{matching to the} full solution \added[id=R3]{\eqref{eq:seriesol} at $y= \pm W$}. This linear `far-field solution'\added[id=R3]{, which is valid when $\vert y-y_c\vert\gg s$,}
is shown in figure~\ref{fig:f2}(c) and clearly gives an excellent fit to the nonlinear data. 


\added[id=R3]{Recalling from \eqref{eq:indpressure} that $p=O(W^{-1})$, t}he linear-far field solution \eqref{eq:farfieldsol} for volume-changing 
perturbations suggests that if the channel is sufficiently wide, so that \replaced[id=R3]{$W \sim O(\epsilon^{-1})$}{$W = O(\epsilon^{-2})$}, then the far-field contact-line displacement could become $O(1)$, violating the small displacement assumption of the linear model. \added[id=R3]{
We therefore need to revisit the far-field quadratic approximation (\ref{eq:farfieldsol})}, which should correspond to the arc of a circle having curvature equivalent to that of a large, flat `pancake' catenoid confined between unperturbed plates having the same mean curvature, \hbox{i.e.} the same mean pressure difference
$\Delta p$ as the meniscus.  
The radius $R_d$ of the circular arc which matches onto the contact line is found by computing a catenoid with the same pressure difference $\Delta p$, as evaluated in Appendix \ref{app:catenoid}.  
\begin{figure}
	\centering
	\includegraphics[width=1.03\textwidth]{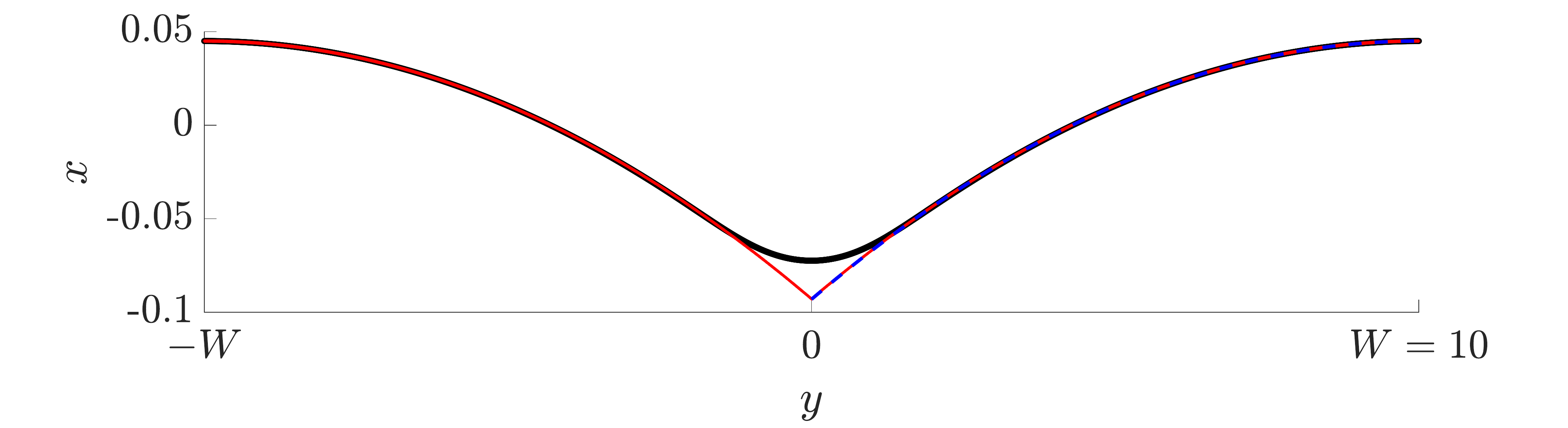}
	\caption{The nonlinear contact-line displacement for a channel of half-width $W=10$, with a contact angle $\phi = 45^\circ$ and perturbations to the channel wall of amplitude $\epsilon = 0.01$ and $s^2=1$. The \replaced[id=R2]{thick black solid line denotes}{black circles denote} the nonlinear upper contact-line data computed in Surface Evolver $x_{\text{cl}}$. The red line denotes the quadratic linear far-field solution \eqref{eq:farfieldsol}, $x_0 + R \cos \added[id=R2]{\tilde \theta} + \epsilon \hat x_p^+(y)$, with $C \approx 0.18$. The blue dashes denote the arc of a circle \added[id=ECJ]{of radius $R_d$, where $R_d$ is determined as part of the solution of the boundary-value problem described in Appendix \ref{app:catenoid} with $\Delta p \approx -1.4167$ found from Surface Evolver}.}
	\label{fig:5}
\end{figure}
Figure~\ref{fig:5} shows how the computed far-field shape of the contact line is captured well by \replaced[id=ECJ]{both the linear far-field solution \eqref{eq:farfieldsol} and by the circular arc computed from the catenoid solution.}{the linear far-field solution \eqref{eq:farfieldsol} and slightly better by the circular arc computed from the catenoid solution.}

In summary, the pressure change induced by the net volume changes of the wall perturbations generates curvature of the contact line away from the perturbations, whereas other geometric features of the perturbations influence contact-line shapes locally.  We therefore expect 
that the same far-field behaviour should exist if the perturbations are not aligned, as we shall test in the next section. 

\subsection{Non-aligned perturbations}
We now consider perturbations which are not aligned, i.e. $y_c^\pm \neq 0$. We specifically consider configurations of perturbations that are sufficiently far apart to be considered as isolated perturbations.

\begin{figure}
	\centering
(a)\includegraphics[width=0.8\textwidth]{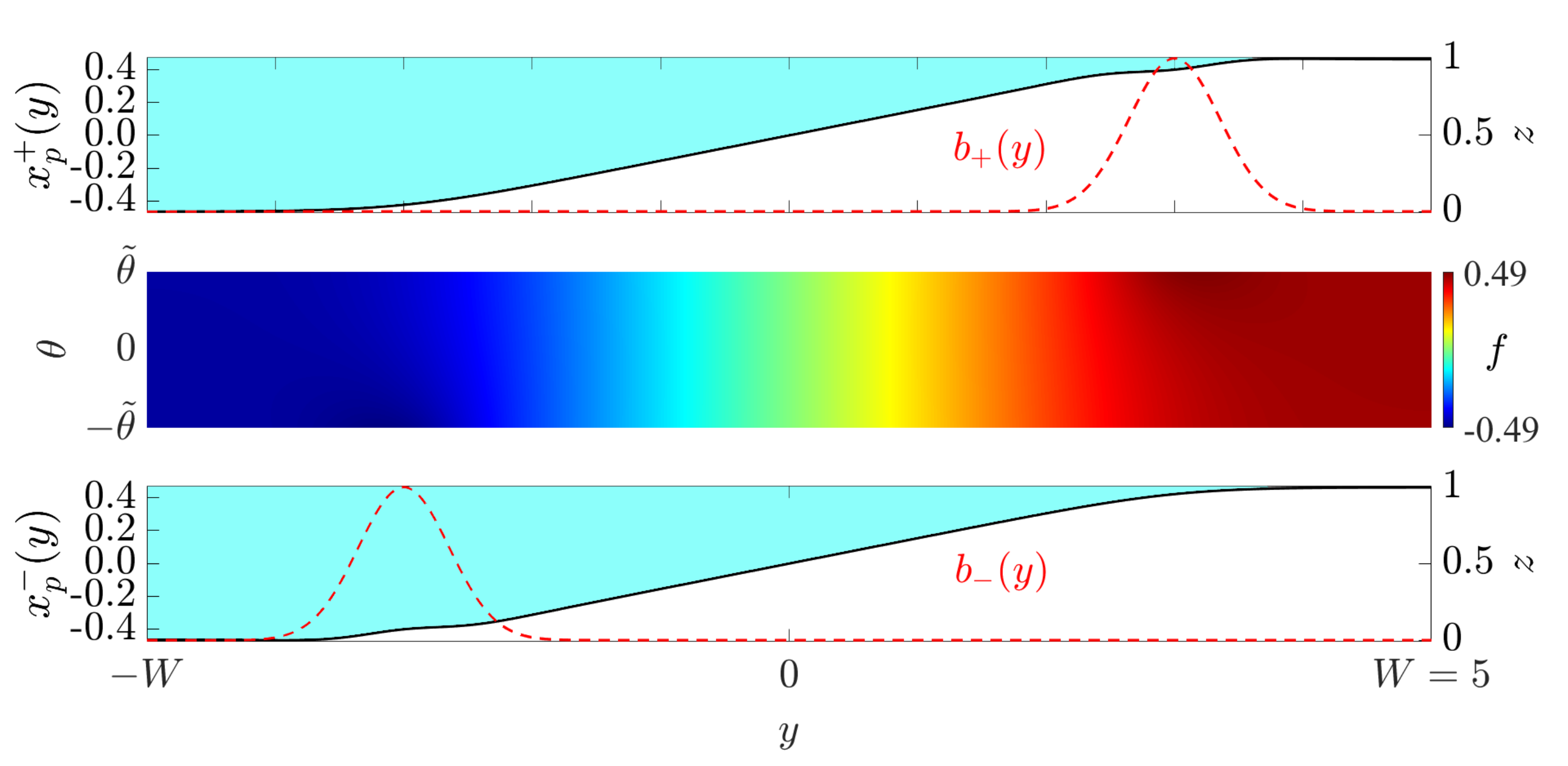}\label{fig:3a}\\(b)\includegraphics[width=0.8\textwidth]{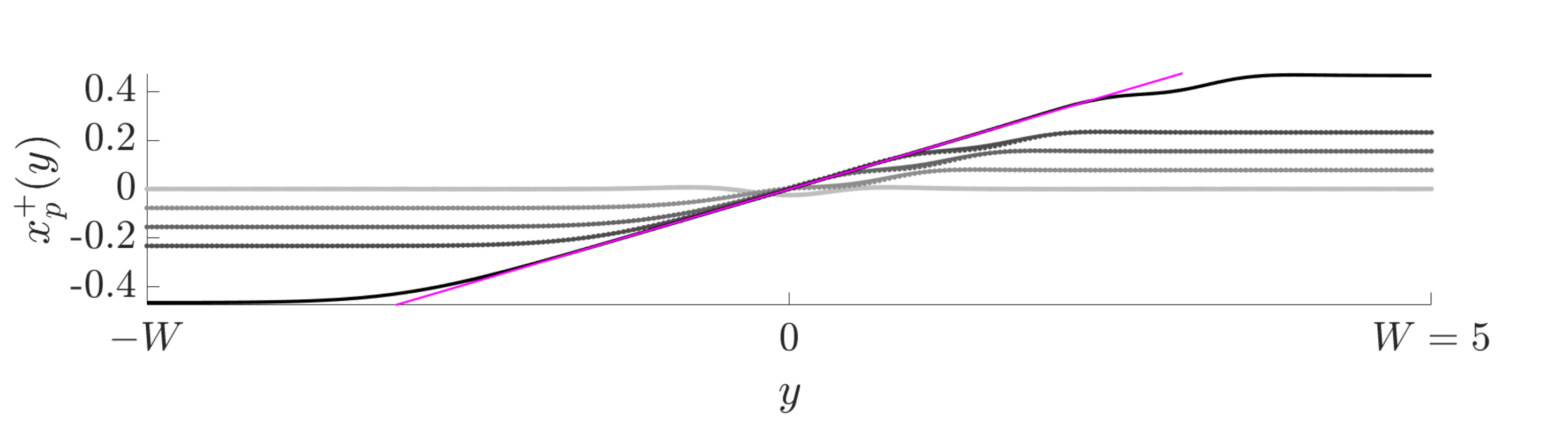}\label{fig:3b}\\(c)\includegraphics[width=0.8\textwidth]{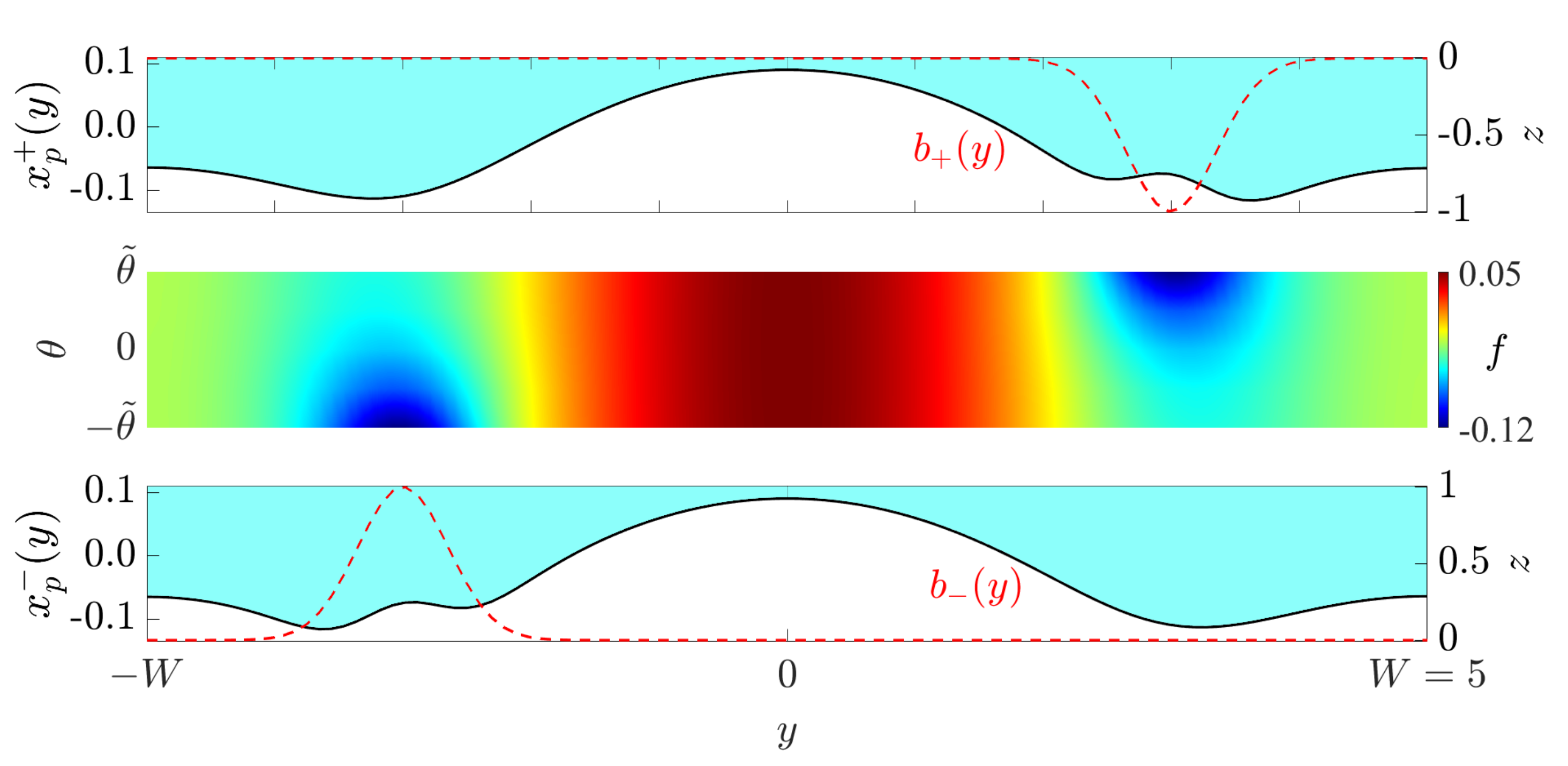}\label{fig:3c}\\(d)\includegraphics[width=0.8\textwidth]{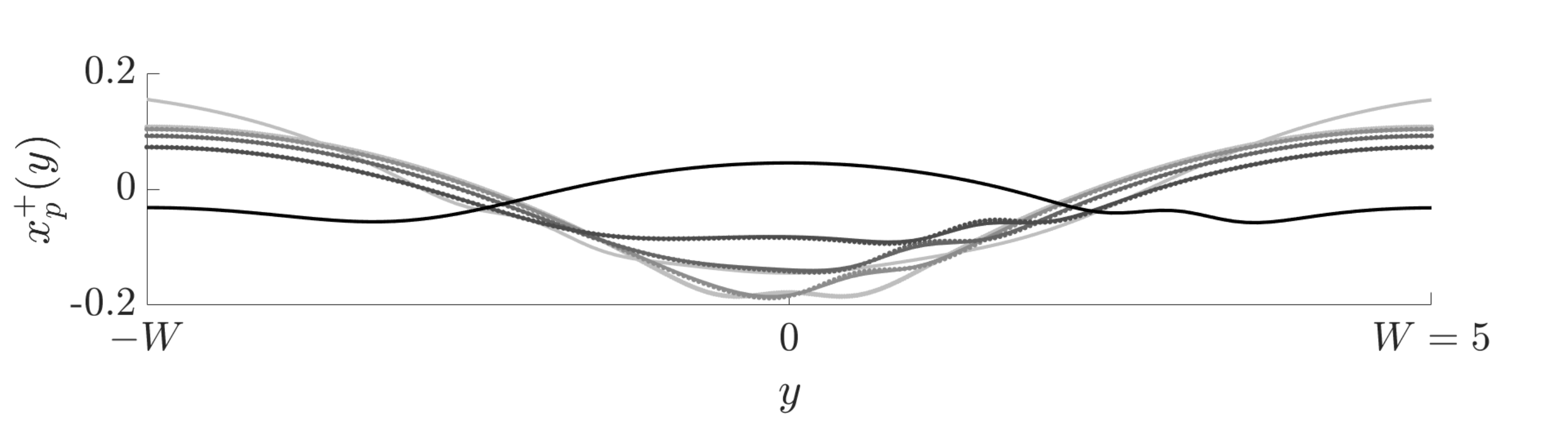}\label{fig:3d}
	\caption{The linear solution $f(y, \theta)$ and upper contact-line displacement for channel-volume-preserving (a, b) and channel-volume-changing (c, d) perturbations, for channel half-width $W = 5$ and contact angle $\phi = 85^{\circ}$. The upper and lower wall perturbations are given by $B_\pm(y) = 0.01 \exp((y - y_c^\pm)^2/0.25)$. \replaced[id=ECJ]{$y_c^\pm=\pm3$}{$y_c=1$} in (a, c), while (b, d) shows contact-line displacement $x_p^+$ for separations varying from $y_c^\pm = \pm 3$ (black) through $\pm(1.5, 1, 0.5)$ to $y_c^\pm = 0$ (light gray). Solid lines denote the displacement calculated via the linear solution, $x_p^+(y)$ from \eqref{eq:cleqn}; circles \added[id=R2]{(thicker lines)} denote the nonlinear displacement $(x_{\text{cl}}-x_0-R\cos \added[id=R2]{\tilde \theta})/\epsilon$ where $x_{\text{cl}}$ is the upper contact-line data computed in Surface Evolver. The pink line in (b) is the line $x = \alpha y$, where the slope $\alpha$ is given in \eqref{eq:alphaeqn}. }
	\label{fig:f3}
\end{figure}

 We compute the non-aligned solutions to the linear model using second-order-accurate central finite differences (Appendix \ref{app:linmethods}), with step sizes of $\Delta y = 0.05$ in the $y$-direction and $\Delta \theta = 0.03$ in the $\theta$ direction. Figure~\ref{fig:f3} shows solutions of the linear problem for perturbations which have been separated so that $b_\pm(y)$ are centred at $\pm y_c$. 
The separation of the channel-volume-preserving perturbations causes the contact line to bend away from the side wall; this \replaced[id=R3]{deflection}{`scattering'} \replaced[id=R3]{of the upper and lower contact lines occurs due to the presence of a perturbation on either the upper or lower wall.}{is induced by the presence of the perturbation on a single wall, and we note that the shape of the contact line is affected by an obstacle on either wall.} Isolated ridges and grooves cause the contact lines to move towards the liquid and vapour phases respectively. 
In contrast, 
channel-volume-changing perturbations (figure~\ref{fig:f3}c, d) induce non-local bending of the contact line in the far-field which can again be described using arcs of circles. 

\added[id=R3]{We wish to understand the deflection mechanism so that we may choose perturbations to engineer specific contact line shapes. }In the channel-volume-preserving case (figure~\ref{fig:f3}a,b), let $\alpha$ be the gradient of the contact-line displacement in the centre of the channel, between the perturbations. For perturbations of sufficiently small amplitude, $\alpha$ is approximately equal to the \replaced[id=R3]{angle of deflection, i.e. the angle}{`scattering angle'} that the contact line makes with the horizontal. We can obtain an approximation for $\alpha$ by considering the solution in the neighbourhood of an isolated ridge or groove: consider the Helmholtz \replaced[id=R3]{equation}{problem} \eqref{eq:helmholtz}, with zero pressure difference $p$ \replaced[id=R3]{(to obtain contact line solutions with uniform gradient). Then the problem for a single isolated ridge or groove perturbation $b_+(y)$ on the upper wall, centred at some $y=y_c$,}{induced by the perturbations, perturbed by a single isolated ridge or groove on the upper wall so that} \added[id=R3]{will admit a solution to the Helmholtz equation with $f(y, \theta) = 0$ for $y_c-y \gg s$ and $f(y, \theta) \approx \alpha (y-y_c) \cos \theta$ for $y-y_c \gg s$.} 
Exploiting the self-adjointness of the Helmholtz operator and boundary conditions using the method given in Appendix~\ref{app:indpressure} with a test function $g(\theta) = \cos \theta$, we obtain 
\begin{equation}\label{eq:alphaeqn}
\alpha \approx \frac{1}{R^2} \frac{1}{\cos \added[id=R2]{\tilde \theta} \sin\added[id=R2]{\tilde \theta} + \added[id=R2]{\tilde \theta} }\int_{-{\infty}}^{\infty} b_+(y) \ \mathrm d y,
\end{equation}
\added[id=R3]{where the integral is taken over the full width of the isolated perturbation $b_+$.}
Thus we anticipate that for Gaussian perturbations, the parameters that most affect the \replaced[id=R3]{deflection}{scattering}will be the volume of the perturbation and the contact angle. While the linear theory allows for an $O(\epsilon)$ contact-line displacement in the $x$-direction, \replaced[id=R3]{the displacement $\epsilon \alpha y$ of the deflected solution can in principle become $O(1)$ in a sufficiently wide channel (if $y-y_c = O(\epsilon^{-1})$); thus the solution can in principle be matched to a straight meniscus for which the $x$-displacement is larger.}{the \replaced[id=R3]{deflected}{scattered} solution can in principle be matched to a straight meniscus for which the $x$-displacement is larger (in a sufficiently wide channel).}  
\begin{figure}
	\hspace{-1.1cm}
	\centering
	\includegraphics[width=1.074\textwidth]{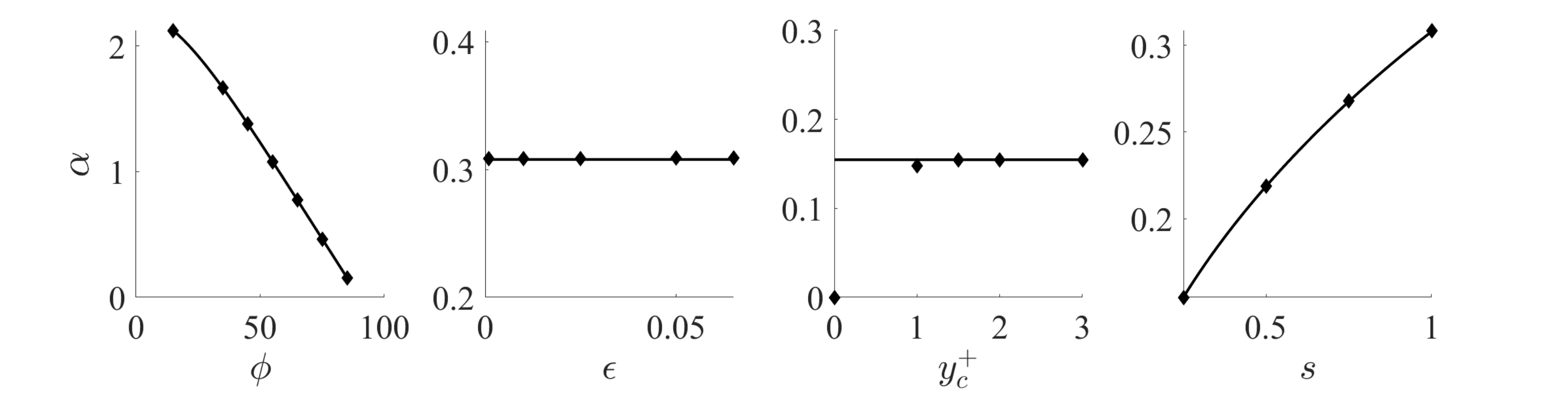}\label{fig:4a}\\ (a) \hspace{2.75cm} (b) \hspace{2.75cm} (c) \hspace{2.75cm} (d)
	\caption{The slopes $\alpha$ of the upper contact-line displacement $x_p^+(y)$ in the centre of the channel for channel-volume-preserving perturbations. (a) Varying contact angle $15^\circ \leq \phi \leq 85^\circ$ with separation $y_c^+ = 3$, perturbation width $s^2 = 0.25$, and perturbation amplitude $\epsilon = 0.01$. (b) Varying amplitude $0.001 \leq \epsilon \leq 0.065$ with $y_c^+ = 3$, $s^2=1$ and $\phi = 85$. (c) Varying separation $0 \leq y_c^\pm \leq 3$ with $s^2=0.25$, $\epsilon = 0.01$ and $\phi = 85$. (d) Varying perturbation width $0.25 \leq s^2 \leq 1$ with $y_c^+ = 3$, $\phi = 85$ and $\epsilon = 0.01$. The solid lines denote the values of $\alpha$ calculated using \eqref{eq:alphaeqn}. The diamonds denote the numerical values of $\alpha$ calculated empirically from the contact-line data.} 
	\label{fig:f4}
\end{figure}
Figure~\ref{fig:f4} shows the values of $\alpha$ found empirically, together with the theoretical prediction \eqref{eq:alphaeqn}, for varying perturbation separation, width, amplitude and contact angle. There is an excellent agreement with the theoretical predictions except for small $y_c^+$, i.e. as long as the perturbations are not too close together; this is expected because it violates the assumption that the perturbations can be treated as isolated ridges and grooves. 

\subsection{Weakly corrugated channels}
Based on the discussion above, we now consider the linear model for channels with a series of small-amplitude ridges and grooves on the upper wall to form weakly corrugated channel walls. Thus we consider perturbations of the form
\begin{equation}\label{eq:gausspertcorrug}
b_\pm(y) = \sum_{k=0}^{K} a_k^{\pm} \mathrm e^{-(y-y_{c_k}^\pm)^2/s^2},
\end{equation}
where $y_{c_k}^\pm$ are the locations of the ridges and grooves on the upper and lower walls and $a_k^{\pm}=\pm1$ depending on whether ridges or grooves are chosen. We assume that the ridges and grooves are sufficiently spaced so that \replaced[id=R3]{we can treat each perturbation as a single isolated ridge or groove that causes deflection of the contact line in the way described above, so that we can predict the deflection angle due to each ridge and groove using \eqref{eq:alphaeqn}.}{the local solution in the region of the perturbation each ridge and groove can be considered in terms of the isolated perturbation problem described above, which allows us to predict the local \replaced[id=R3]{deflection}{scattering} due to each ridge and groove \eqref{eq:alphaeqn}.}
\begin{figure}
	\centering
	(a)\includegraphics[width=1.03\textwidth]{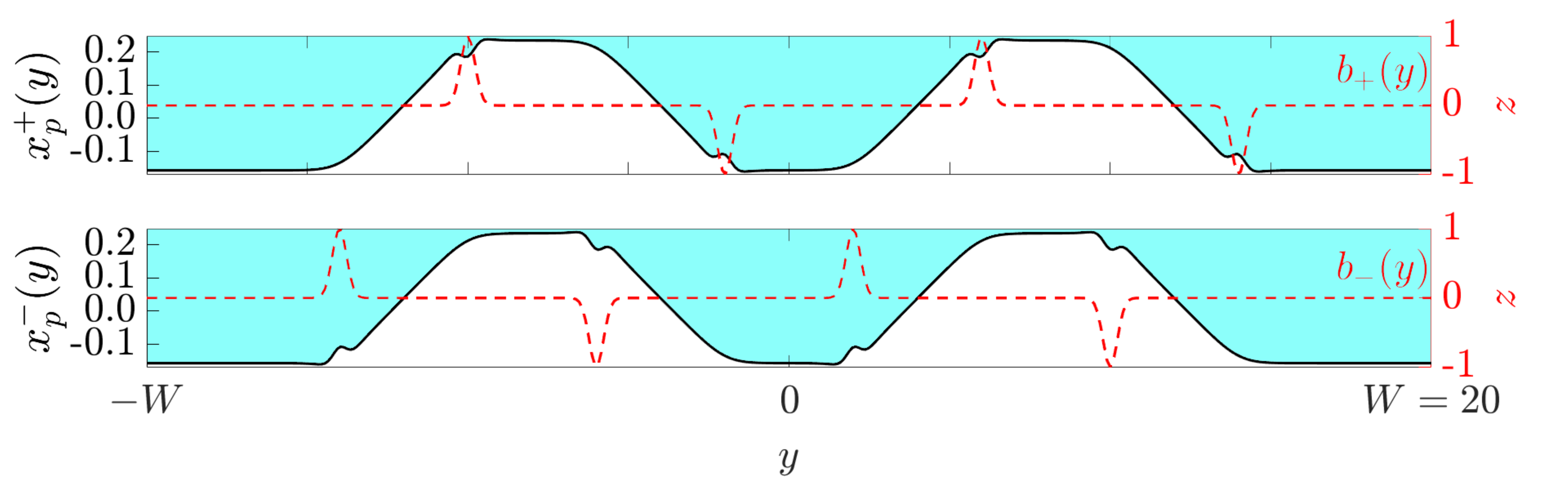}	\label{fig:7}
	(b)\includegraphics[width=1.03\textwidth]{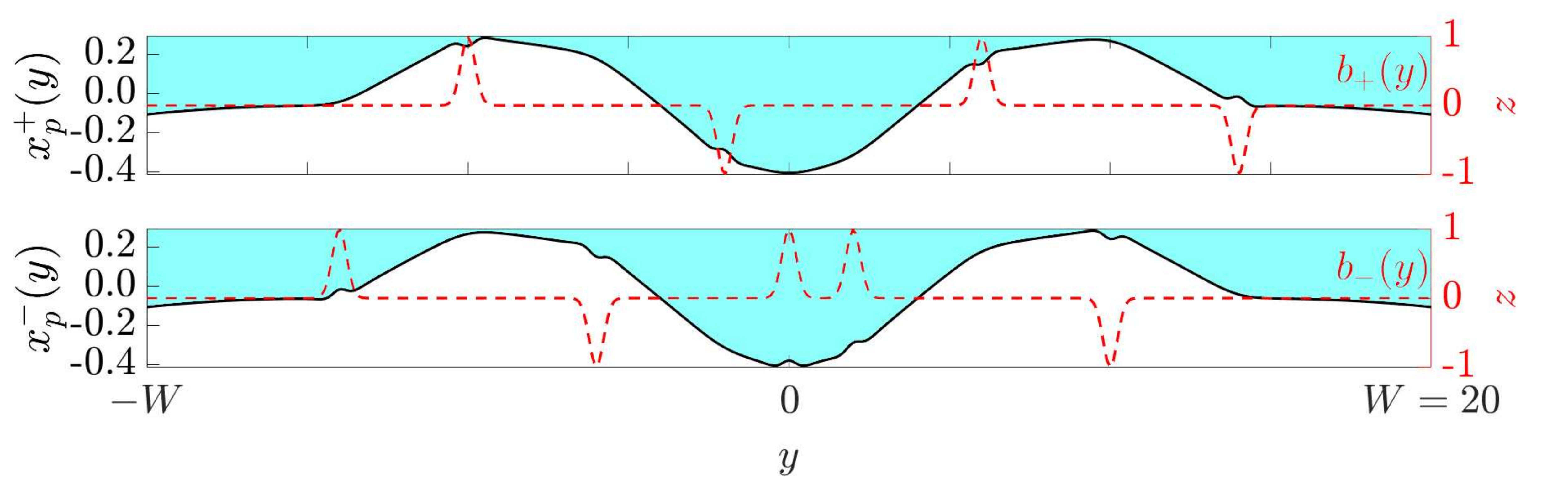}
	\label{fig:7b}
	\caption{The upper and lower contact lines (left axis, solid black line), together with the perturbations (right axis, dashed red line) for $(a)$ a channel-volume-preserving configuration and $(b)$ a channel-volume-changing configuration in a channel of half-width $W=20$. The perturbations are defined by \eqref{eq:gausspertcorrug} with ridges on the lower wall at $y = -14, 2$, grooves on the lower wall at $y=-6, 10$, ridges on the upper wall at $y=-2, 14$ and grooves on the upper wall at $y=-10, 6$. The perturbation in $(b)$ has an extra ridge on the lower wall at $y=0$ to make it channel-volume changing. All perturbations have width $s^2=0.1$ and the contact angle is $\phi = 85^\circ$. }
	\label{fig:8}
\end{figure}
Figure~\ref{fig:8} shows the upper and lower contact-line displacements $x_p^\pm(y)$ for a weakly corrugated channel with alternating ridges and grooves on each wall so that the contact lines take the shape of a letter `M'. The contact-line displacement for the channel-volume-preserving configuration is shown in figure~\ref{fig:8}\added[id=R3]{(a)}; this solution describes a meniscus with zero induced mean curvature, and thus the contact-line displacement is flat in the far-field and has sections of varying slope\replaced[id=R3]{. Because each perturbation can be treated in isolation, the gradient of}{;} each slope is described by \eqref{eq:alphaeqn}. Again, ridges push the contact line towards the liquid phase and grooves allow the contact line to move towards the vapour phase. The contact-line slope varies smoothly, and again the shape of the contact line is affected by an obstacle on either wall so that it takes, for example, a ridge/groove on the lower wall followed by a ridge/groove on the upper wall to reverse the gradient of the contact line. 

In figure~\ref{fig:8}(b) we consider the same series of grooves and ridges and then add an extra ridge on the lower wall at $y=0$ so that the configuration is now channel-volume-changing. Qualitatively, the shape is unchanged but the change in mean curvature is now non-zero. Thus the `sections' of contact line between each ridge and groove are now locally parabolic; each parabola \replaced[id=R2]{is a local approximation to}{can be described as} the arc of a circle which forms the contact line of a catenoid with the same mean curvature \added[id=R2]{as the static} meniscus.

There are, of course, a plethora of possible smoothly-varying contact-line shapes that can be made using channel-volume-changing/preserving configurations, for small-amplitude perturbations (up to approximately 10\% of the channel height). It is possible to specify these shapes \textit{a priori} using just the boundary data using either  \eqref{eq:alphaeqn} for the required gradients in the channel-volume-preserving case, or  \eqref{eq:indpressure} to deduce the pressure of the catenoid with circular contact line that matches onto the parabolas in the channel-volume-changing case, together with the known direction in which the contact line will move for either ridges or grooves. 

\section{Discussion}\label{sec:conclusion}

In this study, we have 
quantified the displacement of the contact line of a static meniscus in a rectangular channel arising from the presence of isolated ridges and grooves imposed on the channel walls.  
We have shown that small-amplitude perturbations which change the channel volume induce a change in the mean curvature of the meniscus, inducing long-range curvature of the contact line, via (\ref{eq:indpressure}). For very wide channels, this curvature matches onto the arc of a catenoid whose radius is found by matching the pressure differences. Meanwhile, small-amplitude isolated non-aligned perturbations which do not change the channel volume generate a contact-line shape that is approximately piecewise linear.  We derived an approximation to the \replaced[id=R3]{deflection}{scattering} angle between adjacent linear segments (\ref{eq:alphaeqn}), showing a dependence on the volume of the groove or ridge.  
This makes it possible in principle to engineer contact-line shapes by choosing the location and order of the ridges and grooves.

We validated predictions of the linearised model against fully nonlinear solutions obtained using Surface Evolver.  However it remains unclear at present how the closed-form results (\ref{eq:indpressure}, \ref{eq:alphaeqn}), derived using the self-adjointness of the Helmholtz equation, might be extended to the nonlinear regime.  While these predictions of the induced pressure and \replaced[id=R3]{deflection}{scattering} angle show dependence on the volume of ridges or grooves, they mask more subtle dependence on the precise shape of the perturbations.  For example, when there is no induced pressure change, the contact-line displacement near a ridge or groove mirrors approximately the curvature of the wall shape (figure~\ref{fig:fig6}a), which is a bounded function for the Gaussian wall perturbations chosen here. Sharper perturbations, having derivatives varying on very short lengthscales, can be expected to lead to dramatically different outcomes, as outlined in Appendix~\ref{app:sharp}. We avoided these extreme cases here by ensuring that $b_{\pm}(y)$ is analytic and not too narrow.

A natural extension of this study is to consider 
perturbations with curvature in two directions (such as isolated bumps).  These too can be expected to generate long-range deflections of the contact line.  However nonlinear effects (associated with large amplitudes or sharp asperities) will likely need to be taken into account in order to capture effects such as contact line hysteresis, arising as the contact line is moved slowly backwards and forwards over the bump. 
Similarly, the approach take\added[id=R3]{n} here could equally be extended to consider the sensitivity of the meniscus to changes in the contact angle arising from coating portions of the channel wall with suitable chemicals. 
In practice, however, a continuous gradient of contact angle may be much more difficult to achieve experimentally than smoothly-varying perturbations, which may appear naturally in an industrial or biological setting.  \added[id=R3]{The present study considered perturbing a `straight' meniscus with zero Gaussian curvature; a further generalisation that merits investigation is to consider a curved base state, to accommodate contact angles at lateral walls that deviate from $\pi/2$.}

The solution structures identified here will support future studies of gas/liquid interfaces moving at low capillary numbers through domains having isolated geometric features, be these engineered in order to achieve a specific outcome or naturally occurring roughness.  We have shown that, even when these features are smooth, isolated and of small amplitude, significant long-range deflections of the meniscus are possible.\\

Declaration of Interests. The authors report no conflict of interest.

\appendix
\section{Linearised problem for zero contact angle}\label{app:zerocont}
When $\phi = 0$ the linearised boundary condition \eqref{eq:lbc2} vanishes, requiring 
expansion up to $O(\epsilon^2)$. We write the interface location as
\begin{subequations}
\begin{align}
r &= R + \epsilon f_1(y, \theta) + \epsilon^2 f_2(y, \theta) + O(\epsilon^3), \\
\theta &\in \left[- \added[id=R2]{\tilde \theta}+ \epsilon \, \Theta_{1-}(y) + \epsilon^2 \Theta_{2-}(y), \ \added[id=R2]{\tilde \theta} +\epsilon \, \Theta_{1+}(y) + \epsilon^2 \Theta_{2+}(y) \right],
\end{align}
\end{subequations}
where $R = \tfrac 1 2$. The pressure difference $\Delta p$ is assumed to be 
$\Delta p = -R^{-1} + \epsilon p_{1} + \epsilon^2 p_{2}+\dots$.
After linearising the Young--Laplace equation \eqref{eq:YoungLap3d}, the $O(\epsilon)$ expression gives the equation for $f_1$,
\begin{equation}
\frac 1 {R^2} f_1 + \frac 1 {R^2} f_{1\theta\theta} + f_{1yy} =p_{1}.
\end{equation}
Similarly, the $O(\epsilon)$ terms in the linearisation of boundary conditions \eqref{eq:nlbc1} and \eqref{eq:nlbc3} give the boundary conditions on $f_1$, 
\begin{align}
f_1\left(y, \pm \frac \pi 2 \right) = b_{\pm}(y) ; \quad f_{1y}(\pm W, \theta) = 0.
\end{align}
The equation relating the change in meniscus shape $f(y, \theta)$, to the contact line location $\Theta_{1 \pm}(y)$, can be found at $O(\epsilon^2)$:
\begin{equation}
f_{1\theta}\left(y, \pm \frac \pi 2\right) = R \ \Theta_{1\pm}(y).
\end{equation}
Therefore we recover exactly the conditions \eqref{eq:lbc1}, \eqref{eq:lbc2} with $\phi =0$.
The volume constraint \eqref{eq:volume} and independent pressure condition \eqref{eq:indpressure} remain the same. 

\section{The pressure in the linear problem}\label{app:indpressure}
For the linearised problem, we can derive an independent equation for the pressure by using the fact that the Helmholtz operator is self-adjoint. Consider the linear problem (\ref{eq:helmholtz})-(\ref{eq:volume}) and a smooth, twice differentiable test function $g(\theta): [-\added[id=R2]{\tilde \theta}, \ \added[id=R2]{\tilde \theta}] \to \mathbb R$, such that
\begin{align}
R^{-2} [ g'' + g] = a \in \mathbb R; \quad g(\pm \added[id=R2]{\tilde \theta}) = \gamma_{\pm}, \quad g'(\pm \added[id=R2]{\tilde \theta}) = \zeta_\pm.
\end{align}
We multiply the Helmholtz equation \eqref{eq:helmholtz} by $g$, and then integrate over the domain $D = [-\added[id=R2]{\added[id=R2]{\tilde \theta}}, \added[id=R2]{\added[id=R2]{\tilde \theta}}] \times [-W, W]$. Then, defining $\tilde \nabla = (\partial_y, R^{-1}\partial_\theta)$, we obtain
\begin{equation}\label{eq:int1}
\int_D a f + \tilde \nabla \cdot (g \tilde \nabla f - f \tilde \nabla g) \ \mathrm d A = \int_D gp \ \mathrm d A. 
\end{equation}
Rewriting the divergence terms on the left hand side as integrals over closed curves, we then integrate and apply the boundary conditions at the side walls, \eqref{eq:lbc3}, and the boundary conditions on $g$ to give
\begin{equation}\label{eq:in2}
\int_D a f \ \mathrm d A +  R^{-2}  \int_{-W}^{W} [- \gamma_- f_\theta(y, -\added[id=R2]{\added[id=R2]{\tilde \theta}}) + \zeta_- f(y, -\added[id=R2]{\added[id=R2]{\tilde \theta}}) + \gamma_+ f_\theta(y, \added[id=R2]{\added[id=R2]{\tilde \theta}}) - \zeta_+ f(y, \added[id=R2]{\added[id=R2]{\tilde \theta}})] \ \mathrm d y = \int_D gp \ \mathrm d A. 
\end{equation}
We now pick a test function $g(\theta) = \cos\theta$ (so that $a=0$) and apply the boundary conditions \eqref{eq:lbc1} on $f$, which leads to an independent equation for the pressure:
\begin{equation}\label{eq:pressure}
p = \frac{1}{4 W R^2 \sin (\added[id=R2]{\added[id=R2]{\tilde \theta}})} \int_{-W}^{W} [ b_+(y) - b_-(y)] \ \mathrm d y.
\end{equation}

\added[id=ECJ]{We also use this method to derive an approximation to the deflection angle $\alpha$ as given in \eqref{eq:alphaeqn}. To derive this relationship, we again use a test function $g(\theta) = \cos \theta$ but we take `far-field' boundary conditions across an isolated wall perturbation of $f_y \rightarrow 0$ for $(y-y_c)/s \rightarrow -\infty$ and $f_y \rightarrow \alpha \cos \theta$ for $(y-y_c)/s \rightarrow \infty$.}
\section{Solving the nonlinear problem with Surface Evolver }\label{app:evolverexpl}
We implement the nonlinear problem using Surface Evolver \citep{evolver} which uses a gradient-descent method to iterate towards a surface of minimum energy. The energy of the triangulated surface is defined as a scalar function of all the vertex coordinates. The iterative process forces the vertices into a configuration which is closer to an energy minimum, subject to any global constraints on the surface or local constraints on the vertices. For the problem outlined in \S\ref{sec:model}, the only force acting on the liquid-vapour interface is surface tension and thus we minimise the surface energy of the meniscus. The constraints are the boundary conditions \eqref{eq:nlbc1}-\eqref{eq:nlbc3} together with the volume constraint.

The boundary conditions on the upper and lower channel walls, \eqref{eq:nlbc2}-\eqref{eq:nlbc3}, are implemented by fixing the energy of the channel walls. We define the (non-dimensional) surface tension due to the presence of the solid wall as $\gamma_S = \gamma_{SL}/\gamma_{LV}-\gamma_{SV}/\gamma_{LV}$, where $\gamma_{SL}$ and $\gamma_{SV}$ are solid-liquid and solid-vapour surface tensions. Then if $\alpha_w$ is the contact angle at the solid-liquid and liquid-vapour interface, by Young's equation, $\gamma_S = -\cos\alpha_w$, and the wall energy is 
\begin{equation}\label{eq:wallen}
E_{\mbox{wall}} = \iint_{\mbox{wall}} -\cos\alpha_w \ \mathrm d S.
\end{equation}
Thus to specify the boundary conditions \eqref{eq:nlbc2}-\eqref{eq:nlbc3} above we fix the wall energies by imposing contact angles $\alpha_w = \phi$ on the upper and lower walls at $z = \pm 1/2 + B_{\pm}(y)$ and $\alpha_w = \pi/2$ on the side walls at $y = \pm W$ (so that the energy of the side walls is zero).

In practice, for computational efficiency, only the liquid-vapour interface is triangulated and refined. Then the wall energy integral \eqref{eq:wallen} for the upper and lower walls is rewritten as a line integral using Stokes Theorem: if $\boldsymbol w = (w_x, w_y, w_z)$ is such that $(\nabla \times \boldsymbol w)\cdot \uvec v_\pm = -\cos(\phi) $ (where again $\uvec v_\pm$ is the unit outward normal to the upper and lower channel walls), then defining $\partial_{\scalebox{.7}{wall}}$ as the boundary of the wall, the wall energy is
\begin{equation}
E_{\mbox{wall}} = \oint_{ \partial_{\scalebox{.7}{wall}}} \boldsymbol w \cdot \ \mathrm d \boldsymbol r.
\end{equation}
For the upper and lower walls described by $z = \pm 1/2 + B_{\pm}(y)$, we can choose, for example, $w_x = w_z = 0$ and $w_y = -x \cos\phi \sqrt{1 + B_\pm(y)^2}$. The integral around the closed curve is implemented internally by Surface Evolver along the edges specified by the user, with their orientation defined such that the unit normal is outward-pointing.

Boundary condition \eqref{eq:nlbc1} is imposed by constraining the contact-line vertices to lie on the upper wall of the channel. This is a local condition on each vertex.

The fixed volume of liquid $V_L$ is handled in Surface Evolver as a global constraint on the possible energy configurations that the surface can take; that is, it removes one degree of freedom from the problem.

The mesh refinement is handled by Surface Evolver using a basic subdivision; we also equiangulate the mesh after each iteration. We converge to an energy minimum using the following process:
\begin{enumerate}
	\item iterate on a fixed mesh until the solution is accurate to a specified tolerance;
	\item refine the mesh and check the difference between the energy on the new mesh and the old mesh. While the difference is greater than a specified tolerance, repeat step (i). 
\end{enumerate}
We use a tolerance of $10^{-6}$ for the accuracy of the solution on each mesh and the energy difference between meshes. We ensure a global minimum has been reached by using a second-order gradient descent method to check for positive eigenvalues near the equilibrium.  
\section{Numerical solution of the linear problem}\label{app:linmethods}
We solve the linear problem (\ref{eq:helmholtz})-(\ref{eq:volume}) with Gaussian boundary data $b_\pm(y) = \pm \exp(-(y-y_c^\pm)^2/s^2)$ in a rectangular domain $-W \leq y \leq W$, $-\added[id=R2]{\added[id=R2]{\tilde \theta}} \leq \theta \leq \added[id=R2]{\added[id=R2]{\tilde \theta}}$. For general $y_c^\pm$, we integrate the Helmholtz equation (\ref{eq:helmholtz}) using second-order-accurate central finite differences with step lengths $\Delta y$ and $\Delta \theta$ in the $y$ and $\theta$ directions respectively. We denote the value of the solution $f$ at $y = k \Delta y$, $\theta = j \Delta k$ by $f_k^j$ for $0 \leq k \leq M+1$, $0 \leq j \leq N+1$, so that $y = W$ is approximated by $(M+1)\Delta y$ and $\theta = \added[id=R2]{\added[id=R2]{\tilde \theta}}$ is approximated by $(N+1)\Delta \theta$. We discretise the Helmholtz equation \eqref{eq:helmholtz} on the interior of the grid as
\begin{align}\label{eq:discHelm}
\frac{1}{\Delta y^2} f_{k+1}^j + \frac{1}{\Delta y^2}  f_{k-1}^j + \frac 1{\Delta \theta^2 R^2}f_k^{j+1} +\frac 1{\Delta \theta^2 R^2} f_k^{j-1} + \left( \frac{1}{R^2} - \frac{2}{\Delta \theta^2 R^2} - \frac{2}{\Delta y^2}\right)f_k^j = p, \nonumber \\ 
(1 \leq k \leq M, \ 1 \leq j \leq N).
\end{align}
We use the boundary conditions \eqref{eq:lbc1}-\eqref{eq:lbc2} to show that
\begin{align}\label{eq:discBC1}
2 \Delta \theta \sin(\theta_{N+1}) f_k^{N+1} - 2 \Delta \theta  b_+(y_k) +(3f^{N+1}_k - 4f^N_k +f^{N-1}_k)  \cos(\theta_{N+1}) = 0,\\
2 \Delta \theta \sin(\theta_{0}) f^{0}_k  + 2 \Delta \theta b_-(y_k) + (3f^{0}_k - 4f^1_k +f^{2}_k)  \cos(\theta_{0}) = 0,
\end{align}
meanwhile the Neumann boundary conditions \eqref{eq:lbc3} give
\begin{equation}\label{eq:discBC2}
-3f_0^j + 4f_1^j -f_2^j = 0,\quad 3f_{M+1}^j - 4f_M^j +f_{M-1}^j \quad \mbox{for} \quad 1 \leq j \leq N.
\end{equation}
We then obtain a system of $(N+1)(M+1)$ equations which we solve subject to the volume constraint \eqref{eq:volume}, which we discretise using Simpson's rule. 

\added[id=R3]{The discretised system is solved using a direct solver which takes advantage of the sparsity in the matrix structure. The grid size is chosen so that the solution of the discretised finite difference system at each grid point is accurate to three decimal places compared to the truncated analytical series solution, which is known at each grid point to high accuracy (truncated terms had size $O(10^{-16})$, see \S\ref{sec:analyticsol}).}
\section{The catenoid problem: governing equations and solution}\label{app:catenoid} 
	\begin{figure}
	\centering
	\includegraphics[width=1\textwidth]{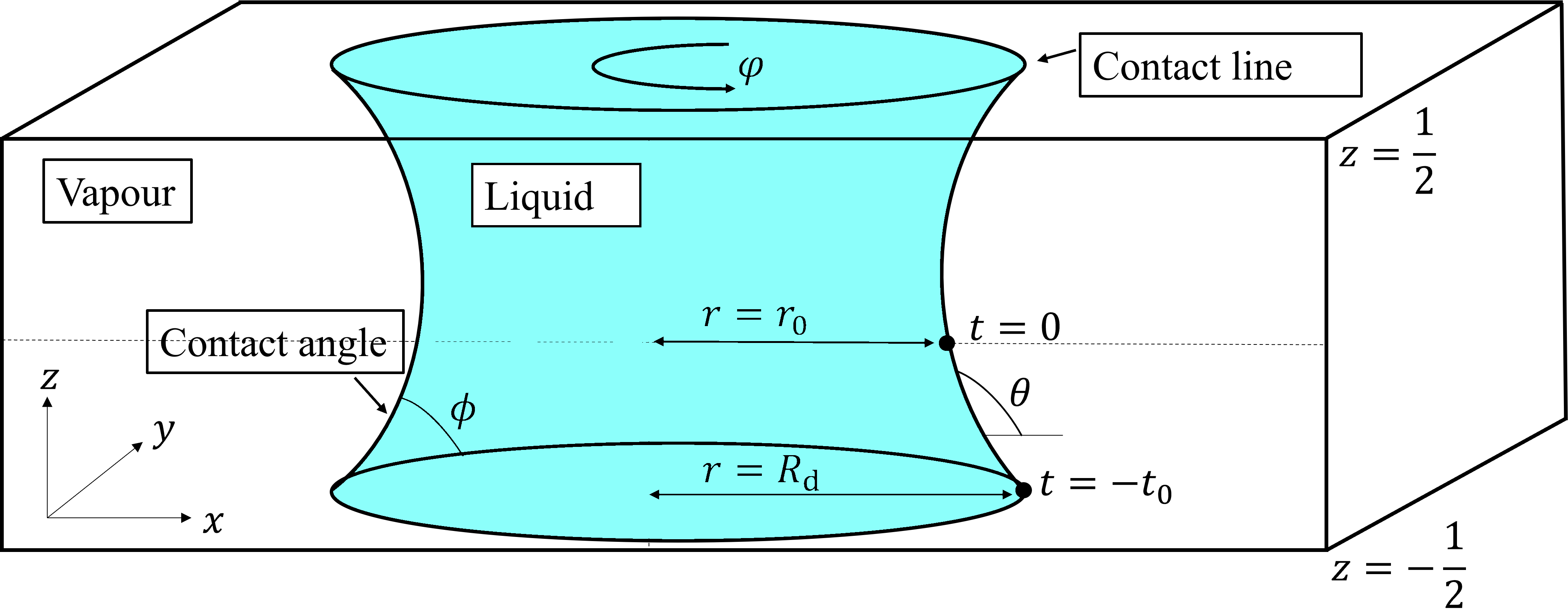}
	\caption{A catenoid in a rectangular channel with height $-\tfrac 1 2 \leq z \leq \tfrac 1 2 $ and solid-liquid contact angle $\phi$\added[id=ECJ]{.} }
	\label{fig:a1}
\end{figure}
Consider a catenoid with solid-liquid contact angle $0 \leq \phi < \pi/2$ in a rectangular channel. \added[id=ECJ]{Note that we use the term `catenoid' here to describe the general shape of the interface as an `inverted droplet', however it need not be a surface of zero mean curvature.} \replaced[id=ECJ]{The catenoid interface}{which is described by cylindrical polar coordinates $(r, \varphi, z)$ as shown in figure~\ref{fig:a1}. The catenoid is axisymmetric with respect to the azimuthal angle $\varphi$ and is also symmetric about $z=0$. Therefore, for a fixed $\varphi = \varphi_0$, the surface of the catenoid} is described by arc-angle coordinates $(r(t), z(t), \theta(t))$ for $-t_0 \leq t \leq t_0$ \added[id=ECJ]{and is axissymmetric with respect to the azimuthal angle $\varphi$ in cylindrical polar coordinates $(r, \varphi, z)$}. We take $t=0$ to be at $z=0$ as shown in figure~\ref{fig:a1} so that $(r(0), z(0), \theta(0) )= (r_0, 0, \pi/2)$. Meanwhile the channel walls are at \replaced[id=R3]{$t = \pm t_0$}{$s=\pm 1$} so that $(r(t_0), z(t_0), \theta(t_0)) = (R_d, 1/2, \phi)$ and $(r(-t_0), z(-t_0), \theta(-t_0)) = (R_d, -1/2, \pi-\phi)$. \replaced[id=ECJ]{The}{Since the} catenoid is symmetric about $z=0$, \added[id=ECJ]{therefore} without loss of generality we can consider the interface from $0\leq t \leq t_0$. Then, $r$ and $z$ depend \added[id=ECJ]{implicitly} on $t$ as 
\begin{align}\label{eq:dropodes1}
\mydiff{ r}{t} &= \cos \theta, \quad \mydiff{ z}{t} = \sin\theta, \quad 0 \leq t \leq t_0, \\
r(0) &= r_0, \ r(t_0) = R_d; \quad z(0) = 0, \ z(t_0) = \frac 1 2.
\end{align}
The unit normal to the interface pointing into the vapour phase at $\varphi = $constant is given by $\uvec {\hat n}  = \sin \theta \uvec{\hat r} -\cos \theta \uvec {\hat z}$. Thus, \replaced[id=R3]{the Young--Laplace equation is}{solving the Young--Laplace equation shows that}
\begin{equation}
    \Delta p = \nabla \cdot \uvec {\hat n} = \cos  \theta \myparI{ \theta}{r} + \frac{\sin  \theta}{  r} + \sin \theta \myparI{\theta}{z} = \mydiff{\theta}{t} + \frac{\sin \theta}{  r}, 
\end{equation}
so that the final equation in the system is
\begin{align}\label{eq:dropodes2}
\mydiff{ \theta}{t} &= \Delta p -\frac {\sin \theta}{r} , \quad 0 \leq t \leq t_0, \\
\theta(0) &= \frac \pi 2, \ \theta(t_0) = \phi.
\end{align}
The ODEs \eqref{eq:dropodes1},\eqref{eq:dropodes2}, together with the boundary conditions\deleted[id=ECJ]{described above}, form a boundary value problem for the arc-angle components $ r, z, \theta$. 

A large-radius asymptotic solution ($r_0\gg 1$) to this system can be found by writing
\begin{align}
    r(t) &= r_0 + r_1(t) + \frac{r_2(t)}{r_0} + \cdots, \quad \theta(t) = \theta_0(t) + \frac{\theta_1(t)}{r_0} + \cdots, \\
        z(t) &= z_0(t) + \frac{z_1(t)}{r_0} + \cdots, \quad \Delta p = (\Delta p)_0 + \frac{(\Delta p)_1}{r_0} + \cdots, \\
        t_0 &= L_0 + \frac 1 {r_0} L_1 + \cdots.
\end{align}
Solving the leading order problem, we find from the leading-order approximation ($\theta_0=(\Delta p)_0t+\pi/2$, $z_0=(\sin(\Delta p)_0 t))/(\Delta p)_0$) that the pressure difference across the catenoid is
\begin{equation}
    (\Delta p)_0 = -2 \cos \phi,
\end{equation}
which is consistent with the pressure difference of the unperturbed static liquid-vapour meniscus in the rectangular channel\added[id=ECJ]{, while the leading-order approximation to the catenoid radius and the endpoint of the curve $t_0$ is}
\added[id=ECJ]{\begin{equation}
   r_1(t) = \frac{\cos((\Delta p)_0 t)}{(\Delta p)_0} - \frac 1 {(\Delta p)_0}, \quad L_0 = \frac{2 \phi - \pi}{2}.
\end{equation}}
Solving the $O(r_0^{-1})$ problem, we find that
\begin{equation}\label{eq:asymppres}
    (\Delta p)_1 ={\frac {\sin \left( \tilde \theta \right) \cos \left( \tilde \theta \right) +\tilde \theta}{2\sin \left( \tilde \theta \right) }}.
\end{equation}
Thus,
\added[id=ECJ]{eliminating $r_0$ from truncated expansions for $r(t_0)$ and $\Delta p$, the expression
\begin{equation}
    R_d \approx \frac{(\Delta p)_1}{\Delta p - (\Delta p)_0 } + r_1(L_0)
    \label{eq:largeRd}
\end{equation}
}
gives the large-radius approximation for the catenoid for any given $\Delta p$. 

\begin{figure}
\centering
	\includegraphics[width=0.7\textwidth]{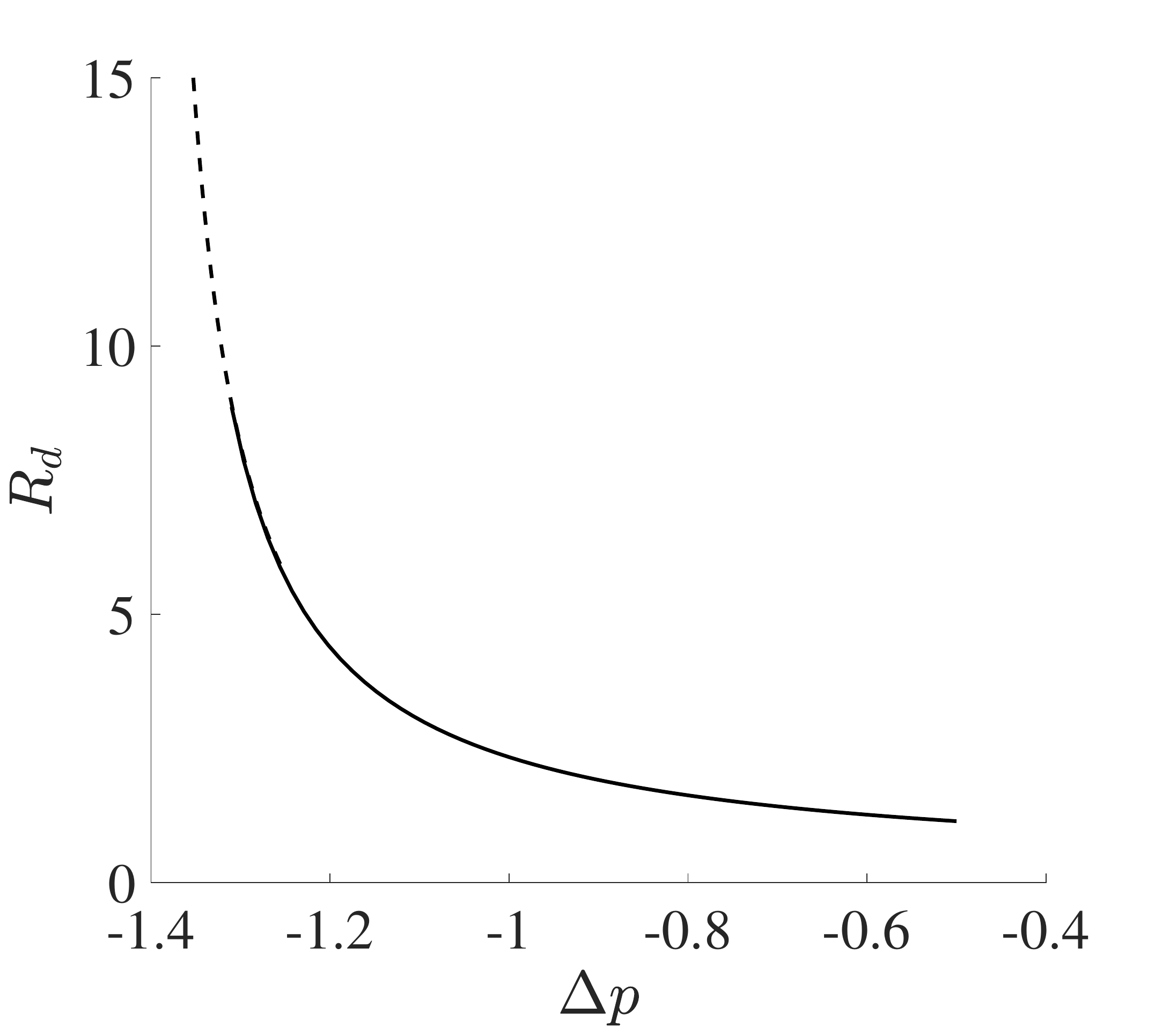}
	\caption{The maximum radius \replaced[id=R3]{$R_d$}{$R_2$} of a catenoid for varying pressure difference $\Delta p$, for a contact angle $\phi = 45^\circ$. The solid line denotes the numerical solution for $R_d$. \added[id=ECJ]{The dashed line denotes the asymptotic solution (\ref{eq:largeRd}).} The location of the asymptote is at $\Delta p = -2\cos(\pi/4) \approx -1.414$. }
	\label{fig:a2}
\end{figure}

We can \added[id=R3]{also} solve for catenoids with smaller radii numerically. First, we eliminate $t$ to obtain \replaced[id=R3]{a nonlinear boundary-value problem where the unknown radius $R_d$ is to be determined as part of the solution for given $\Delta p$ and $\phi$:}{an eigenvalue problem for the unknown radius $R_d$:}
\begin{subequations}
\begin{align}
\mydiff{ \theta}{ z} = \frac{\Delta p}{\sin \theta} - \frac{1}{  r}, &\quad \mydiff{ r}{ z}  = \frac{\cos  \theta}{\sin  \theta}, \\
 r \left(\frac 1 2 \right)  =  R_d, \quad  \theta\left(0 \right) &= \frac \pi 2, \quad  \theta\left(\frac 1 2 \right)	= \phi. 
\end{align}
\end{subequations}
We solve this problem numerically using the \textsc{Matlab} routine `bvp4c' \citep{kierzenka2001bvp} and thus for any static meniscus with a given pressure difference \added[id=R3]{(mean curvature)} and contact angle, we can find the \deleted[id=R3]{corresponding contact-line catenoid} radius \replaced[id=ECJ]{$R_d$}{$ R_s$} \replaced[id=R3]{of the circular contact line of the catenoid which has the same pressure difference (mean curvature). The relationship between pressure difference and catenoid radius}{which matches onto the far-field static meniscus solution; this relationship} is shown in figure~\ref{fig:a2}.  \added[id=ECJ]{It matches closely the asymptote (\ref{eq:largeRd}) for $R_d\gg 1$.}

\added[id=R3]{The large-radius catenoid solution describes the curved meniscus shape (\ref{eq:farfieldsol}) far from the wall perturbation, as we now demonstrate.}
The circular contact line is locally described by a parabola\added[id=R3]{. To see this, take a catenoid with contact line of radius $R_d$ centred on $x=x_0$, $y=W$. Its contact line lies along 
$\left(x-x_0\right)^2+(y-W)^2=R_d^2$.
Because the solution is translationally invariant, $x_0$ may be chosen such that the catenoid passes through $x=0$, $y=y_c$. When $W\ll R_d$ and the centre of the catenoid lies in $x>0$, we may describe the base of the contact line using
\begin{subequations}
    \label{eq:parabolic}
\begin{align}
    x&=x_0-\sqrt{R_d^2-(W-y)^2} \\
    &\approx x_0 - R_d\left(1-\frac{(W-y)^2}{2R_d^2}+\dots \right) \approx C_0 + \frac{1}{2R_d}(W-y)^2,
\end{align}
\end{subequations}
with $C_0$ a constant. Therefore the contact line displacement can be matched to \eqref{eq:farfieldsol} by choosing 
\begin{equation}
    \epsilon p = \frac{\cos \tilde{\theta} \sin\tilde{\theta}+\tilde{\theta}}{2 R_d \sin\tilde{\theta}}=\frac{(\Delta p)_1}{R_d},
    \label{eq:eprd}
\end{equation}
and therefore approximates the contact line in $y_c<y\leq W$ when $\vert y-y_c \vert \gg s$.  This pressure-radius relationship is consistent with the leading-order relationship found via the asymptotic expansion \eqref{eq:asymppres}, with $r_0 \approx R_d$.
}

\added[id=R3]{We can then assess the limit $W\sim \epsilon^{-1}$, for which the linearisation approximation of \S\ref{sec:modellin} formally breaks down.  Recall that the contact line displacement $\epsilon x_p^\pm$ is $O(\epsilon pW^2)$ with $p=O(1/W)$ (from (\ref{eq:indpressure})), so that the contact line displacement is $O(1)$ and $p=O(\epsilon)$ for $W\sim \epsilon^{-1}$.  However the contact line retains a radius of curvature that is large compared to $W$ ($R_d=O(1/\epsilon^2)$, from (\ref{eq:largeRd}) with $\Delta p - (\Delta p)_0=\epsilon p$), allowing the parabolic approximation (\ref{eq:parabolic}) to be used.  Thus the parabolic description (\ref{eq:farfieldsol}) remains appropriate in this limit (figure~\ref{fig:5}), because of the structure of the catenoid solution.  In contrast, larger-amplitude wall perturbations will cause $R_d$ to fall towards the size of $W$, pushing the contact line towards a more circular shape away from perturbations.}

\section{Sharp ridges and grooves}
\label{app:sharp}

It is well known that a sharp wedge or groove can drive large contact-line displacements \citep{concus1969behavior}, and we have so far restricted attention to Gaussian perturbations (\ref{eq:gausspert}) having width $s$ no smaller than $O(\epsilon^{1/2})$, where $\epsilon$ is the wall-perturbation amplitude.

\begin{figure}
	(a)\includegraphics[width=1.1\textwidth]{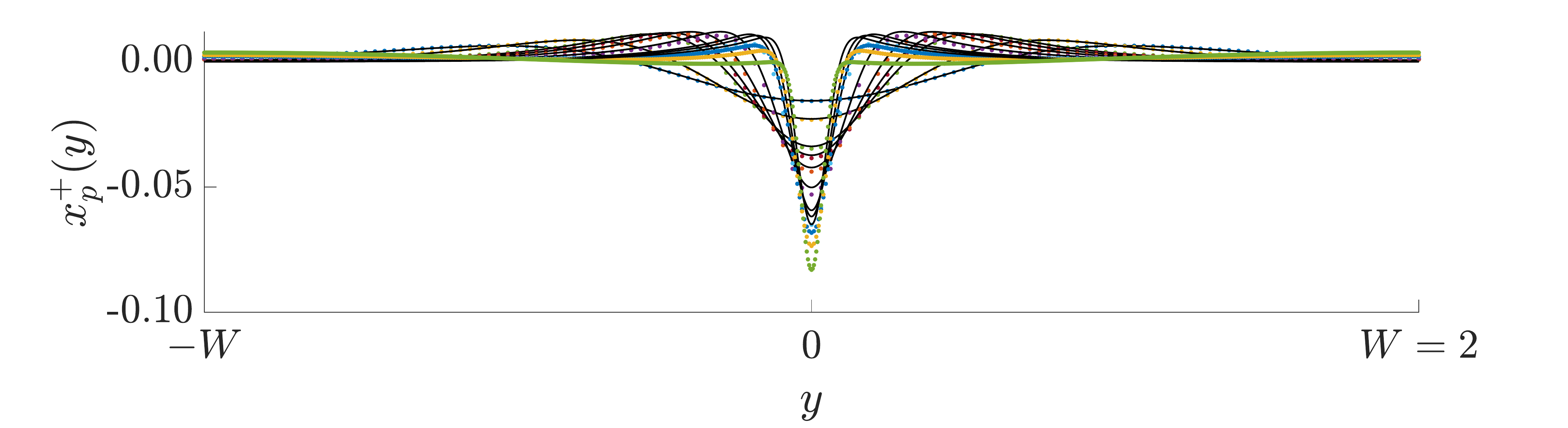}
	(b)\includegraphics[width=1.1\textwidth]{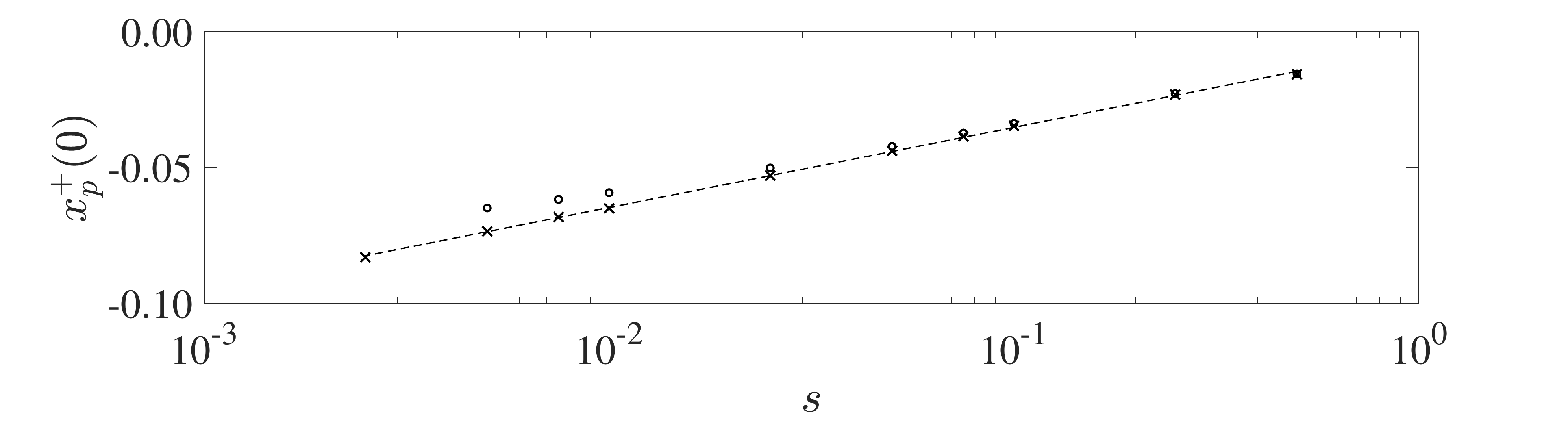}
	\caption{(a) The upper contact line displacement $\hat x_p^+(y)$ for a channel-volume-preserving Gaussian perturbation $B_\pm(y)=0.01\exp(-y^2/s^2)$ with $s^2$ taking values $0.0025, 0.005, 0.0075, 0.01, 0.025, 0.05, 0.075, 0.1, 0.25$ and $0.5$. The black lines denote the linear solution, computed using the series solution \eqref{eq:seriesol}, while the coloured dots denote the Surface Evolver solution. The contact angle is $\phi = 85^\circ$ and the channel half-width is $W=2$. (b) The upper contact line displacement at $y=0$, $\hat x_p^+(0)$, with a logarithmic scale on the $x$-axis (values of $s$). The crosses denote the value of $\hat x_p^+(0)$ computed using the Surface Evolver solution; the circles are the corresponding value for the linear solution. The dashed line is $x_p^+(0)=0.0128 \log s – 0.0057$. }
	\label{fig:appf1}
\end{figure}

\added[id=R2]{We now empirically examine what happens to the contact line solution for the Gaussian perturbations $b_\pm(y) = \exp(-y^2/s^2)$ with $s \to 0$.  Figure~\ref{fig:appf1}(a) shows the upper contact line solutions for a channel-volume-preserving perturbation with decreasing $s^2$, with the narrowest computed perturbation having $s^2=0.0025$ in a channel of half-width $W=2$.  Linear solutions computed using the series solution \eqref{eq:seriesol} are compared to Surface Evolver solutions, which are converged to an accuracy of $10^{-8}$ using the process outlined in Appendix \ref{app:evolverexpl}. The amplitude of the contact line displacement increases as the perturbations become narrower; the linear model under-predicts the nonlinear Surface Evolver solution, indicating that nonlinear effects become important as the perturbations become sharper, particularly once $s^2$ approaches $\epsilon=0.01$. We also note that the far field solution does not quite have zero curvature (as the linear model predicts for a channel-volume-preserving perturbation); this again is a nonlinear effect.  Plotting the maximum displacement of the contact line at $y=0$ (figure \ref{fig:appf1}b) shows that the amplitude of the displacement scales like $\log(1/s)$, suggesting blowup may be possible even for analytic boundary forcing.}

\added[id=R2]{A more extreme response can be expected for smaller contact angles and less smooth forcing.}  Consider the case in which the ridge or groove has small amplitude and narrower width (not necessarily Gaussian, but effectively satisfying $s^2\ll \epsilon$).  More specifically, setting $s$ to zero, suppose the lower wall shape ($b_-$), say, has a discontinuity in a derivative at $y=0$, such that $b_-(y)=0$ for $y<0$ and $b_-(y)=y^\gamma$ for $y>0$.  Then $\gamma=0$ corresponds to a step in $b_-$, $\gamma=1$ a corner (a discontinuity in slope) and $\gamma=2$ a jump in the curvature of the wall.  The linearised curvature of the nearby gas-liquid interface, described in general by the Helmholtz equation (\ref{eq:helmholtz}), can be expected to be approximated in the neighbourhood of the discontinuity by $\nabla^2 f\approx 0$.  In the fully wetting case, for example, with $\added[id=R2]{\tilde \theta}=\pi/2$, $b_-$ imposes $f$ along the wall ($\theta=-\pi/2$) via (\ref{eq:lbc1}) and the wall normal derivative $f_n$ determines the contact-line displacement via (\ref{eq:cleqn}).  Introduce polar coordinates $(\varrho,\vartheta)$ centred on $y=0$, such that $\vartheta=0$ ($\vartheta=\pi$) lies along the wall for $y>0$ ($y<0$), and consider first the case of a step ($\gamma=0$). Then $f(\varrho,\vartheta)=-(1-\vartheta/\pi)$ provides a local solution to Laplace’s equation subject to the forcing condition $b_-(y)=H(y)$, where $H$ is a Heaviside function. The corresponding wall normal derivative $f_n$ is then proportional to $1/y$, indicating that the contact line will be displaced in opposite directions on either side of a step.  Further cases follow by integrating with respect to $y$, so that $f_n\propto \log\vert y\vert$ for $\gamma=1$ (the contact line will be displaced along the axis of a corner) and $f_n\propto y\log\vert y\vert-y$ for $\gamma=2$.  These approximate solutions suggest that a very sharp step or a corner in wall shape, even if smoothed over a very short length-scale $s$, will cause substantial deflection of the contact line (violating the linearisation approximation), while a jump in wall curvature will bend the contact line sufficiently for it to have infinite slope with respect to $y$, while remaining continuous.  

In summary, \added[id=R2]{and as indicated by figure~\ref{fig:appf1}}, nonlinear effects will have a leading-order role close to the ridge or groove whenever the wall shape is sufficiently sharp. 

\bibliographystyle{jfm}
\bibliography{staticsbib}

\begin{thebibliography}{65}
\expandafter\ifx\csname natexlab\endcsname\relax\def\natexlab#1{#1}\fi
\def\au#1{#1} \def\ed#1{#1} \def\yr#1{#1}\def\at#1{#1}\def\jt#1{\textit{#1}}
  \def\bt#1{#1}\def\bvol#1{\textbf{#1}} \def\vol#1{#1} \def\pg#1{#1}
  \def\publ#1{#1}\def\arxiv#1{#1}\def\org#1{#1}\def\st#1{\textit{#1}}

\bibitem[Ajaev {\em et~al.\/}(2020)Ajaev, Gatapova \&
  Kabov]{ajaev2020interaction}
{\sc \au{Ajaev, Vladimir~S}, \au{Gatapova, Elizaveta~Ya} \& \au{Kabov, Oleg~A}}
  \yr{2020}  \at{Interaction of advancing contact lines with defects on heated
  substrates}.  \jt{Physical Review E}  \bvol{101}~(2),  \pg{022801}.

\bibitem[Ajaev \& Homsy(2006)]{ajaev2006modeling}
{\sc \au{Ajaev, Vladimir~S} \& \au{Homsy, GM}} \yr{2006}  \at{Modeling shapes
  and dynamics of confined bubbles}.  \jt{Annu. Rev. Fluid Mech.}  \bvol{38},
  \pg{277--307}.

\bibitem[Anna(2016)]{anna2016droplets}
{\sc \au{Anna, Shelley~Lynn}} \yr{2016}  \at{Droplets and bubbles in
  microfluidic devices}.  \jt{Annual Review of Fluid Mechanics}  \bvol{48},
  \pg{285--309}.

\bibitem[Ben-Jacob {\em et~al.\/}(1985)Ben-Jacob, Godbey, Goldenfeld, Koplik,
  Levine, Mueller \& Sander]{ben1985experimental}
{\sc \au{Ben-Jacob, E}, \au{Godbey, R}, \au{Goldenfeld, ND}, \au{Koplik, J},
  \au{Levine, H}, \au{Mueller, T} \& \au{Sander, LM}} \yr{1985}
  \at{Experimental demonstration of the role of anisotropy in interfacial
  pattern formation}.  \jt{Phys. Rev. Lett.}  \bvol{55}~(12),  \pg{1315}.

\bibitem[Bhushan(2019)]{bhushan2019bioinspired}
{\sc \au{Bhushan, B}} \yr{2019}  \at{Bioinspired water collection methods to
  supplement water supply}.  \jt{Phil. Trans. Roy. Soc. A}  \bvol{377}~(2150),
  \pg{20190119}.

\bibitem[Boruvka \& Neumann(1978)]{boruvka1978analytical}
{\sc \au{Boruvka, L} \& \au{Neumann, AW}} \yr{1978}  \at{An analytical solution
  of the {L}aplace equation for the shape of liquid surfaces near a stripwise
  heterogeneous wall}.  \jt{J. Colloid Interface Sci.}  \bvol{65}~(2),
  \pg{315--330}.

\bibitem[Bourges-Monnier \& Shanahan(1995)]{bourges1995influence}
{\sc \au{Bourges-Monnier, C} \& \au{Shanahan, MER}} \yr{1995}  \at{Influence of
  evaporation on contact angle}.  \jt{Langmuir}  \bvol{11}~(7),
  \pg{2820--2829}.

\bibitem[Brakke(1992)]{evolver}
{\sc \au{Brakke, Kenneth~A.}} \yr{1992}  \at{The {S}urface {E}volver}.
  \jt{Experimental Mathematics}  \bvol{1}~(2),  \pg{141--165}.

\bibitem[Brown \& Bhushan(2016)]{brown2016bioinspired}
{\sc \au{Brown, PS} \& \au{Bhushan, B}} \yr{2016}  \at{Bioinspired materials
  for water supply and management: water collection, water purification and
  separation of water from oil}.  \jt{Phil. Trans. Roy. Soc. A}
  \bvol{374}~(2073),  \pg{20160135}.

\bibitem[Calver {\em et~al.\/}(2020)Calver, Gaffney, Walsh, Durham \&
  Oliver]{calver2020thin}
{\sc \au{Calver, S.~N.}, \au{Gaffney, E.~A.}, \au{Walsh, E.~J.}, \au{Durham,
  W.~M.} \& \au{Oliver, J.~M.}} \yr{2020}  \at{On the thin-film asymptotics of
  surface tension driven microfluidics}.  \jt{J. Fluid Mech.}  \bvol{901}.

\bibitem[Cassie(1948)]{cassie1948contact}
{\sc \au{Cassie, A. B.~D.}} \yr{1948}  \at{Contact angles}.  \jt{Discuss.
  Faraday Soc.}  \bvol{3},  \pg{11--16}.

\bibitem[Cassie \& Baxter(1944)]{cassie1944wettability}
{\sc \au{Cassie, A. B.~D.} \& \au{Baxter, S.}} \yr{1944}  \at{Wettability of
  porous surfaces}.  \jt{Trans. Faraday Soc.}  \bvol{40},  \pg{546--551}.

\bibitem[Chung {\em et~al.\/}(2008)Chung, Park, Shin, Lee \&
  Kwon]{chung2008guided}
{\sc \au{Chung, Su~Eun}, \au{Park, Wook}, \au{Shin, Sunghwan}, \au{Lee,
  Seung~Ah} \& \au{Kwon, Sunghoon}} \yr{2008}  \at{Guided and fluidic
  self-assembly of microstructures using railed microfluidic channels}.
  \jt{Nature materials}  \bvol{7}~(7),  \pg{581--587}.

\bibitem[Comanns {\em et~al.\/}(2015)Comanns, Buchberger, Buchsbaum,
  Baumgartner, Kogler, Bauer \& Baumgartner]{comanns2015directional}
{\sc \au{Comanns, P.}, \au{Buchberger, G.}, \au{Buchsbaum, A.},
  \au{Baumgartner, R.}, \au{Kogler, A.}, \au{Bauer, S.} \& \au{Baumgartner,
  W.}} \yr{2015}  \at{Directional, passive liquid transport: the texas horned
  lizard as a model for a biomimetic}.  \jt{J. Roy. Soc. Interface}
  \bvol{12}~(109),  \pg{20150415}.

\bibitem[Concus \& Finn(1969)]{concus1969behavior}
{\sc \au{Concus, P} \& \au{Finn, R}} \yr{1969}  \at{On the behavior of a
  capillary surface in a wedge}.  \jt{Proc. Nat. Acad. Sci. USA}
  \bvol{63}~(2),  \pg{292}.

\bibitem[Cox(1983)]{cox1983spreading}
{\sc \au{Cox, R.~G.}} \yr{1983}  \at{The spreading of a liquid on a rough solid
  surface}.  \jt{J. Fluid Mech.}  \bvol{131},  \pg{1--26}.

\bibitem[Dorsey \& Martin(1987)]{dorsey_martin_1987}
{\sc \au{Dorsey, A.~T.} \& \au{Martin, O.}} \yr{1987}  \at{{S}affman--{T}aylor
  fingers with anisotropic surface tension}.  \jt{Phys. Rev. A}  \bvol{35},
  \pg{3989--3992}.

\bibitem[Dussan~V(1979)]{dussan1979spreading}
{\sc \au{Dussan~V, EB}} \yr{1979}  \at{On the spreading of liquids on solid
  surfaces: static and dynamic contact lines}.  \jt{Ann. Rev. Fluid Mech.}
  \bvol{11}~(1),  \pg{371--400}.

\bibitem[Emami {\em et~al.\/}(2013)Emami, Hemeda, Amrei, Luzar, Gad-el Hak \&
  Vahedi~Tafreshi]{emami2013predicting}
{\sc \au{Emami, B}, \au{Hemeda, AA}, \au{Amrei, MM}, \au{Luzar, A}, \au{Gad-el
  Hak, M} \& \au{Vahedi~Tafreshi, H}} \yr{2013}  \at{Predicting longevity of
  submerged superhydrophobic surfaces with parallel grooves}.  \jt{Physics of
  Fluids}  \bvol{25}~(6),  \pg{062108}.

\bibitem[Eral {\em et~al.\/}(2013)Eral, 't~Mannetje \& Oh]{eral2013contact}
{\sc \au{Eral, HB}, \au{'t~Mannetje, DJCM} \& \au{Oh, JM}} \yr{2013}
  \at{Contact angle hysteresis: a review of fundamentals and applications}.
  \jt{Colloid Polymer Sci.}  \bvol{291}~(2),  \pg{247--260}.

\bibitem[Esp{\'\i}n \& Kumar(2015)]{espin2015droplet}
{\sc \au{Esp{\'\i}n, Leonardo} \& \au{Kumar, Satish}} \yr{2015}  \at{Droplet
  spreading and absorption on rough, permeable substrates}.  \jt{Journal of
  Fluid Mechanics}  \bvol{784},  \pg{465--486}.

\bibitem[Fowkes \& Hood(1998)]{fowkes1998surface}
{\sc \au{Fowkes, ND} \& \au{Hood, MJ}} \yr{1998}  \at{Surface tension effects
  in a wedge}.  \jt{Quart J. Mech. Appl. Math.}  \bvol{51}~(4),  \pg{553--561}.

\bibitem[Franco-G{\'o}mez {\em et~al.\/}(2016)Franco-G{\'o}mez, Thompson, Hazel
  \& Juel]{franco2016sensitivity}
{\sc \au{Franco-G{\'o}mez, A}, \au{Thompson, AB}, \au{Hazel, AL} \& \au{Juel,
  A}} \yr{2016}  \at{Sensitivity of {S}affman--{T}aylor fingers to
  channel-depth perturbations}.  \jt{J. Fluid Mech.}  \bvol{794},
  \pg{343--368}.

\bibitem[Franco-G{\'o}mez {\em et~al.\/}(2018)Franco-G{\'o}mez, Thompson, Hazel
  \& Juel]{franco2018bubble}
{\sc \au{Franco-G{\'o}mez, A}, \au{Thompson, AB}, \au{Hazel, AL} \& \au{Juel,
  A}} \yr{2018}  \at{Bubble propagation in {H}ele-{S}haw channels with centred
  constrictions}.  \jt{Fluid Dyn. Res.}  \bvol{50}~(2),  \pg{021403}.

\bibitem[Gao \& McCarthy(2006)]{gao2006contact}
{\sc \au{Gao, L} \& \au{McCarthy, TJ}} \yr{2006}  \at{Contact angle hysteresis
  explained}.  \jt{Langmuir}  \bvol{22}~(14),  \pg{6234--6237}.

\bibitem[Hazel {\em et~al.\/}(2013)Hazel, Pailha, Cox \&
  Juel]{hazel2013multiple}
{\sc \au{Hazel, AL}, \au{Pailha, M}, \au{Cox, SJ} \& \au{Juel, A}} \yr{2013}
  \at{Multiple states of finger propagation in partially occluded tubes}.
  \jt{Phys. Fluids}  \bvol{25}~(6),  \pg{021702}.

\bibitem[He {\em et~al.\/}(2017)He, Yang, Qin, Wen \& Zhang]{he2017roles}
{\sc \au{He, B}, \au{Yang, S}, \au{Qin, Z}, \au{Wen, B} \& \au{Zhang, C}}
  \yr{2017}  \at{The roles of wettability and surface tension in droplet
  formation during inkjet printing}.  \jt{Scientific Reports}  \bvol{7}~(1),
  \pg{1--7}.

\bibitem[Hong \& Family(1988)]{hong1988bubbles}
{\sc \au{Hong, DC} \& \au{Family, F}} \yr{1988}  \at{Bubbles in the
  {H}ele-{S}haw cell: Pattern selection and tip perturbations}.  \jt{Phys. Rev.
  A}  \bvol{38}~(10),  \pg{5253}.

\bibitem[Howell~Jr {\em et~al.\/}(2005)Howell~Jr, Mott, Fertig, Kaplan, Golden,
  Oran \& Ligler]{howell2005microfluidic}
{\sc \au{Howell~Jr, Peter~B}, \au{Mott, David~R}, \au{Fertig, Stephanie},
  \au{Kaplan, Carolyn~R}, \au{Golden, Joel~P}, \au{Oran, Elaine~S} \&
  \au{Ligler, Frances~S}} \yr{2005}  \at{A microfluidic mixer with grooves
  placed on the top and bottom of the channel}.  \jt{Lab on a Chip}
  \bvol{5}~(5),  \pg{524--530}.

\bibitem[Huh \& Mason(1977)]{huh1977roughness}
{\sc \au{Huh, C.} \& \au{Mason, S.G.}} \yr{1977}  \at{Effects of surface
  roughness on wetting (theoretical)}.  \jt{J. Colloid Interface Sci.}
  \bvol{60}~(1),  \pg{11--38}.

\bibitem[Jansons(1985)]{jansons1985moving}
{\sc \au{Jansons, KM}} \yr{1985}  \at{Moving contact lines on a two-dimensional
  rough surface}.  \jt{J. Fluid Mech.}  \bvol{154},  \pg{1--28}.

\bibitem[Johnson \& Dettre(1964)]{johnson1964contact}
{\sc \au{Johnson, RE} \& \au{Dettre, RH}} \yr{1964}  \at{Contact angle
  hysteresis. iii. study of an idealized heterogeneous surface}.  \jt{J. Phys.
  Chem.}  \bvol{68}~(7),  \pg{1744--1750}.

\bibitem[Ju {\em et~al.\/}(2012)Ju, Bai, Zheng, Zhao, Fang \&
  Jiang]{ju2012multi}
{\sc \au{Ju, J}, \au{Bai, H}, \au{Zheng, Y}, \au{Zhao, T}, \au{Fang, R} \&
  \au{Jiang, L}} \yr{2012}  \at{A multi-structural and multi-functional
  integrated fog collection system in cactus}.  \jt{Nature Commun.}
  \bvol{3}~(1),  \pg{1--6}.

\bibitem[Khabiry {\em et~al.\/}(2009)Khabiry, Chung, Hancock, Soundararajan,
  Du, Cropek, Lee \& Khademhosseini]{khabiry2009cell}
{\sc \au{Khabiry, Masoud}, \au{Chung, Bong~Geun}, \au{Hancock, Matthew~J},
  \au{Soundararajan, Harish~Chandra}, \au{Du, Yanan}, \au{Cropek, Donald},
  \au{Lee, Won~Gu} \& \au{Khademhosseini, Ali}} \yr{2009}  \at{Cell docking in
  double grooves in a microfluidic channel}.  \jt{Small}  \bvol{5}~(10),
  \pg{1186--1194}.

\bibitem[Khademhosseini {\em et~al.\/}(2005)Khademhosseini, Yeh, Eng, Karp,
  Kaji, Borenstein, Farokhzad \& Langer]{khademhosseini2005cell}
{\sc \au{Khademhosseini, Ali}, \au{Yeh, Judy}, \au{Eng, George}, \au{Karp,
  Jeffrey}, \au{Kaji, Hirokazu}, \au{Borenstein, Jeffrey}, \au{Farokhzad,
  Omid~C} \& \au{Langer, Robert}} \yr{2005}  \at{Cell docking inside microwells
  within reversibly sealed microfluidic channels for fabricating multiphenotype
  cell arrays}.  \jt{Lab on a Chip}  \bvol{5}~(12),  \pg{1380--1386}.

\bibitem[Khademhosseini {\em et~al.\/}(2004)Khademhosseini, Yeh, Jon, Eng, Suh,
  Burdick \& Langer]{khademhosseini2004molded}
{\sc \au{Khademhosseini, Ali}, \au{Yeh, Judy}, \au{Jon, Sangyong}, \au{Eng,
  George}, \au{Suh, Kahp~Y}, \au{Burdick, Jason~A} \& \au{Langer, Robert}}
  \yr{2004}  \at{Molded polyethylene glycol microstructures for capturing cells
  within microfluidic channels}.  \jt{Lab on a Chip}  \bvol{4}~(5),
  \pg{425--430}.

\bibitem[Kierzenka \& Shampine(2001)]{kierzenka2001bvp}
{\sc \au{Kierzenka, J.} \& \au{Shampine, L~F}} \yr{2001}  \at{A bvp solver
  based on residual control and the maltab pse}.  \jt{ACM .~Math.~Software}
  \bvol{27}~(3),  \pg{299--316}.

\bibitem[King {\em et~al.\/}(1999)King, Ockendon \& Ockendon]{king1999laplace}
{\sc \au{King, JR}, \au{Ockendon, JR} \& \au{Ockendon, H}} \yr{1999}  \at{The
  {L}aplace--{Y}oung equation near a corner}.  \jt{Quart. J. Mech. Appl. Math.}
   \bvol{52}~(1),  \pg{73--97}.

\bibitem[Lee {\em et~al.\/}(2007)Lee, Hung \& Lee]{lee2007artificial}
{\sc \au{Lee, Philip~J}, \au{Hung, Paul~J} \& \au{Lee, Luke~P}} \yr{2007}
  \at{An artificial liver sinusoid with a microfluidic endothelial-like barrier
  for primary hepatocyte culture}.  \jt{Biotechnology and bioengineering}
  \bvol{97}~(5),  \pg{1340--1346}.

\bibitem[Li {\em et~al.\/}(2017)Li, Zhou, Li, Che, Yao, McHale, Chaudhury \&
  Wang]{li2017topological}
{\sc \au{Li, J}, \au{Zhou, X}, \au{Li, J}, \au{Che, L}, \au{Yao, J},
  \au{McHale, G}, \au{Chaudhury, MK} \& \au{Wang, Z}} \yr{2017}
  \at{Topological liquid diode}.  \jt{Science Adv.}  \bvol{3}~(10),
  \pg{eaao3530}.

\bibitem[Manbachi {\em et~al.\/}(2008)Manbachi, Shrivastava, Cioffi, Chung,
  Moretti, Demirci, Yliperttula \&
  Khademhosseini]{manbachi2008microcirculation}
{\sc \au{Manbachi, Amir}, \au{Shrivastava, Shamit}, \au{Cioffi, Margherita},
  \au{Chung, Bong~Geun}, \au{Moretti, Matteo}, \au{Demirci, Utkan},
  \au{Yliperttula, Marjo} \& \au{Khademhosseini, Ali}} \yr{2008}
  \at{Microcirculation within grooved substrates regulates cell positioning and
  cell docking inside microfluidic channels}.  \jt{Lab on a Chip}
  \bvol{8}~(5),  \pg{747--754}.

\bibitem[Maxworthy(1986)]{maxworthy1986bubble}
{\sc \au{Maxworthy, T}} \yr{1986}  \at{Bubble formation, motion and interaction
  in a {H}ele--{S}haw cell}.  \jt{J. Fluid Mech.}  \bvol{173},  \pg{95--114}.

\bibitem[Mobasseri {\em et~al.\/}(2015)Mobasseri, Faroni, Minogue, Downes,
  Terenghi \& Reid]{mobasseri2015polymer}
{\sc \au{Mobasseri, Atefeh}, \au{Faroni, Alessandro}, \au{Minogue, Ben~M},
  \au{Downes, Sandra}, \au{Terenghi, Giorgio} \& \au{Reid, Adam~J}} \yr{2015}
  \at{Polymer scaffolds with preferential parallel grooves enhance nerve
  regeneration}.  \jt{Tissue Engineering Part A}  \bvol{21}~(5-6),
  \pg{1152--1162}.

\bibitem[Park \& Kumar(2017)]{park2017droplet}
{\sc \au{Park, Joonsik} \& \au{Kumar, Satish}} \yr{2017}  \at{Droplet sliding
  on an inclined substrate with a topographical defect}.  \jt{Langmuir}
  \bvol{33}~(29),  \pg{7352--7363}.

\bibitem[Park {\em et~al.\/}(2006)Park, Toner, Yarmush \&
  Tilles]{park2006microchannel}
{\sc \au{Park, Jaesung}, \au{Toner, Mehmet}, \au{Yarmush, Martin~L} \&
  \au{Tilles, Arno~W}} \yr{2006}  \at{Microchannel bioreactors for
  bioartificial liver support}.  \jt{Microfluidics and nanofluidics}
  \bvol{2}~(6),  \pg{525--535}.

\bibitem[Passerone(2011)]{passerone2011twenty}
{\sc \au{Passerone, A}} \yr{2011}  \at{Twenty years of surface tension
  measurements in space}.  \jt{Micrograv. Sci. Tech.}  \bvol{23}~(2),
  \pg{101--111}.

\bibitem[Pham \& Kumar(2017)]{pham2017drying}
{\sc \au{Pham, Truong} \& \au{Kumar, Satish}} \yr{2017}  \at{Drying of droplets
  of colloidal suspensions on rough substrates}.  \jt{Langmuir}
  \bvol{33}~(38),  \pg{10061--10076}.

\bibitem[Pham \& Kumar(2019)]{pham2019imbibition}
{\sc \au{Pham, Truong} \& \au{Kumar, Satish}} \yr{2019}  \at{Imbibition and
  evaporation of droplets of colloidal suspensions on permeable substrates}.
  \jt{Physical Review Fluids}  \bvol{4}~(3),  \pg{034004}.

\bibitem[Picknett \& Bexon(1977)]{picknett1977evaporation}
{\sc \au{Picknett, RG} \& \au{Bexon, R}} \yr{1977}  \at{The evaporation of
  sessile or pendant drops in still air}.  \jt{J. Colloid Interface Sci.}
  \bvol{61}~(2),  \pg{336--350}.

\bibitem[Prakash {\em et~al.\/}(2008)Prakash, Qu{\'e}r{\'e} \&
  Bush]{prakash2008surface}
{\sc \au{Prakash, M}, \au{Qu{\'e}r{\'e}, D} \& \au{Bush, John~WM}} \yr{2008}
  \at{Surface tension transport of prey by feeding shorebirds: the capillary
  ratchet}.  \jt{Science}  \bvol{320}~(5878),  \pg{931--934}.

\bibitem[Reyssat(2014)]{reyssat2014drops}
{\sc \au{Reyssat, E}} \yr{2014}  \at{Drops and bubbles in wedges}.  \jt{J.
  Fluid Mech.}  \bvol{748},  \pg{641--662}.

\bibitem[Shanahan(1995)]{shanahan1995simple}
{\sc \au{Shanahan, MER}} \yr{1995}  \at{Simple theory of" stick-slip" wetting
  hysteresis}.  \jt{Langmuir}  \bvol{11}~(3),  \pg{1041--1043}.

\bibitem[Stauber {\em et~al.\/}(2014)Stauber, Wilson, Duffy \&
  Sefiane]{stauber2014lifetimes}
{\sc \au{Stauber, JM}, \au{Wilson, SK}, \au{Duffy, BR} \& \au{Sefiane, K}}
  \yr{2014}  \at{On the lifetimes of evaporating droplets}.  \jt{J. Fluid
  Mech.}  \bvol{744}.

\bibitem[Stone {\em et~al.\/}(2004)Stone, Stroock \&
  Ajdari]{stone2004engineering}
{\sc \au{Stone, Howard~A}, \au{Stroock, Abraham~D} \& \au{Ajdari, Armand}}
  \yr{2004}  \at{Engineering flows in small devices: microfluidics toward a
  lab-on-a-chip}.  \jt{Annu. Rev. Fluid Mech.}  \bvol{36},  \pg{381--411}.

\bibitem[Tabeling {\em et~al.\/}(1987)Tabeling, Zocchi \&
  Libchaber]{tabeling1987experimental}
{\sc \au{Tabeling, P}, \au{Zocchi, G} \& \au{Libchaber, A}} \yr{1987}  \at{An
  experimental study of the {S}affman--{T}aylor instability}.  \jt{J. Fluid
  Mech.}  \bvol{177},  \pg{67--82}.

\bibitem[Thompson {\em et~al.\/}(2014)Thompson, Juel \&
  Hazel]{thompson2014multiple}
{\sc \au{Thompson, AB}, \au{Juel, A} \& \au{Hazel, AL}} \yr{2014}  \at{Multiple
  finger propagation modes in {H}ele-{S}haw channels of variable depth}.
  \jt{J. Fluid Mech.}  \bvol{746},  \pg{123--164}.

\bibitem[Weislogel \& Lichter(1998)]{weislogel1996capillary}
{\sc \au{Weislogel, MM} \& \au{Lichter, S}} \yr{1998}  \at{Capillary flow in an
  interior corner}.  \jt{J. Fluid Mech.}  \bvol{373},  \pg{349--378}.

\bibitem[Wenzel(1936)]{wenzel1936resistance}
{\sc \au{Wenzel, RN}} \yr{1936}  \at{Resistance of solid surfaces to wetting by
  water}.  \jt{Ind. Eng. Chem.}  \bvol{28}~(8),  \pg{988--994}.

\bibitem[Xing {\em et~al.\/}(2014)Xing, Deng, Lian, Zhang, Zhang \&
  Zhao]{xing2014multiple}
{\sc \au{Xing, Youqiang}, \au{Deng, Jianxin}, \au{Lian, Yunsong}, \au{Zhang,
  Kedong}, \au{Zhang, Guodong} \& \au{Zhao, Jun}} \yr{2014}  \at{Multiple
  nanoscale parallel grooves formed on si3n4/tic ceramic by femtosecond pulsed
  laser}.  \jt{Applied surface science}  \bvol{289},  \pg{62--71}.

\bibitem[Xu \& Jensen(2017)]{xu2017}
{\sc \au{Xu, F} \& \au{Jensen, OE}} \yr{2017}  \at{Trapping and displacement of
  liquid collars and plugs in rough-walled tubes}.  \jt{Physical Rev. Fluids}
  \bvol{2}~(9),  \pg{094004}.

\bibitem[Xu {\em et~al.\/}(2016)Xu, Lin, Zhang, Shi \& Zheng]{xu2016high}
{\sc \au{Xu, T}, \au{Lin, Y}, \au{Zhang, M}, \au{Shi, W} \& \au{Zheng, Y}}
  \yr{2016}  \at{High-efficiency fog collector: water unidirectional transport
  on heterogeneous rough conical wires}.  \jt{ACS Nano}  \bvol{10}~(12),
  \pg{10681--10688}.

\bibitem[Yang {\em et~al.\/}(2005)Yang, Yang \& Hong]{yang2005droplet}
{\sc \au{Yang, A-S}, \au{Yang, J-C} \& \au{Hong, M-C}} \yr{2005}  \at{Droplet
  ejection study of a picojet printhead}.  \jt{J. Micromech. Microeng.}
  \bvol{16}~(1),  \pg{180}.

\bibitem[Yoshimitsu {\em et~al.\/}(2002)Yoshimitsu, Nakajima, Watanabe \&
  Hashimoto]{yoshimitsu2002effects}
{\sc \au{Yoshimitsu, Zen}, \au{Nakajima, Akira}, \au{Watanabe, Toshiya} \&
  \au{Hashimoto, Kazuhito}} \yr{2002}  \at{Effects of surface structure on the
  hydrophobicity and sliding behavior of water droplets}.  \jt{Langmuir}
  \bvol{18}~(15),  \pg{5818--5822}.

\bibitem[Zheng {\em et~al.\/}(2010)Zheng, Bai, Huang, Tian, Nie, Zhao, Zhai \&
  Jiang]{zheng2010directional}
{\sc \au{Zheng, Y}, \au{Bai, H}, \au{Huang, Z}, \au{Tian, X}, \au{Nie, F-Q},
  \au{Zhao, Y}, \au{Zhai, J} \& \au{Jiang, L}} \yr{2010}  \at{Directional water
  collection on wetted spider silk}.  \jt{Nature}  \bvol{463}~(7281),
  \pg{640--643}.

\bibitem[Zhou {\em et~al.\/}(2012)Zhou, Khodakov, Ellis \&
  Voelcker]{zhou2012surface}
{\sc \au{Zhou, J}, \au{Khodakov, DA}, \au{Ellis, AV} \& \au{Voelcker, NH}}
  \yr{2012}  \at{Surface modification for pdms-based microfluidic devices}.
  \jt{Electrophoresis}  \bvol{33}~(1),  \pg{89--104}.

\end{thebibliography}

\end{document}